\documentclass[usenatbib]{aa}  

\usepackage{graphicx}
\usepackage{multirow}
\usepackage{rotating}
\usepackage{colortbl}
\usepackage{color}
\usepackage[fleqn]{amsmath}
\usepackage{amssymb}
\usepackage{amsfonts}
\usepackage{verbatim}
\usepackage{scalefnt}
\usepackage[percent]{overpic}
\usepackage{ulem}
\usepackage{hyperref}
\usepackage{physics}
\usepackage{url}
\usepackage{placeins}
\usepackage{afterpage}
\usepackage{makecell}

\usepackage{float}
\usepackage{txfonts}

\hypersetup{
  colorlinks=true,        % false: boxed links; true: colored links 
  linkcolor=blue,         % color of internal links                                
  citecolor=blue,         % color of links to bibliography                           
  }

\newcommand{\kh}{\boldsymbol{k_{\rm h}}}
\newcommand{\expo}{e^{i\boldsymbol{k_{\rm h}}\cdot\boldsymbol{x}}}
\newcommand{\expoh}{e^{ik_{\rm h}h}}
\newcommand{\suma}{\sum_{n = -\infty}^{\infty}}
\newcommand{\efou}{e^{i(nK+k_z)z}}
\newcommand{\efouv}{e^{i(n+k_z)z}}

\usepackage{color}
\usepackage{xcolor}

\newcommand{\OK}[1]{\textcolor{red}{[OK]}}

\usepackage{comment}

\begin{document}

 \title{An effective model for magnetic field amplification by the magnetorotational and parasitic instabilities}

   \author{Miquel Miravet-Tenés\thanks{E-mail: m.miravet-tenes@soton.ac.uk}
          \inst{1,2,3} \and Martin E.~Pessah\inst{3}
          }
          
   \institute{Departament d'Astronomia i Astrofísica, Universitat de València, Dr Moliner 50, 46100, Burjassot (València), Spain
     \and 
        Mathematical Sciences and STAG Research Centre, University of Southampton, Southampton SO17 1BJ, UK
    \and
        Niels Bohr International Academy, Niels Bohr Institute, Blegdamsvej 17, DK-2100 Copenhagen Ø, Denmark }

    \date{\today}
    \abstract
    {
  The magnetorotational instability (MRI) is considered a leading mechanism for driving angular momentum transport in differentially rotating astrophysical flows, including accretion disks and protoneutron stars. This process is mediated by the exponential amplification of the magnetic field whose final amplitude is envisioned to be limited by secondary (parasitic) instabilities. In this paper, we investigated the saturation of the MRI via parasitic modes relaxing previous approximations. We carried out the first systematic analysis of the evolution of parasitic modes as they feed off the exponentially growing MRI while being advected by the background shear flow. We provide the most accurate calculation of the amplification factor to which the MRI can grow before the fastest parasitic modes reach a comparable amplitude. We find that this amplification factor is remarkably robust, depending only logarithmically on the initial amplitude of the parasitic modes, in reasonable agreement with numerical simulations. Based on these insights, and guided by numerical simulations, we provide a simple analytical expression for the amplification of magnetic fields responsible for MRI-driven angular momentum transport. Our effective model for magnetic field amplification may enable going beyond the standard prescription for viscous transport currently employed in numerical simulations when the MRI cannot be explicitly resolved.
    }

   \titlerunning{Magnetic field amplification by the MRI}
   \maketitle

\section{Introduction}
\label{sec::intro}

Understanding the mechanism driving angular momentum transport in turbulent magnetized disks is key to moving beyond models based on enhanced viscosity, as originally introduced by~\cite{Shakura:1973} and~\cite{Lynden-Bell:1974}. The inherent differential rotation present in astrophysical disks has the potential to produce unstable magnetic fields with a wide range of properties~\citep{Velokhov:1959,Chandrasekhar:1960, Balbus:1991, Balbus:1992, Pessah:2005,Johansen:2008, Pessah:2012, Das:2018, Mamatsashvili:2020, Squire:2024, Brughmans:2024}.
Ionized rotating fluids with angular frequency profiles decreasing outwards are particularly  prone to developing the so-called magnetorotational instability (MRI) when threaded by a weak magnetic field in the direction perpendicular to the shear~\citep{Balbus:1991}. The turbulence driven by the MRI is considered a leading mechanism enabling accretion in astrophysical disks around compact objects~\citep{Balbus:1998}.

The conditions for the onset of the MRI can be fulfilled in other important astrophysical scenarios. For example, protoneutron stars (PNSs) that result from the core collapse of rotating massive stars can possess regions where the MRI grows faster than the explosion timescale~\citep{Akiyama:2003,Obergaulinger:2006b,Cerda:2008,Rembiasz:2016a,Reboul-Salze:2021}. The MRI can also develop in the binary neutron star (BNS) postmerger phase~\citep{Duez:2006a,Duez:2006b,Siegel:2013,Kiuchi:2018,Kiuchi:2024,Fernandez:2019,Held:2022,Held:2024}. Depending on the total mass of the binary, the system may go through the formation of a short-lived postmerger object, a so-called hypermassive neutron star. This object eventually collapses to a black hole once the support against gravity by rotation or neutrino pressure lessens. During this phase, the MRI can lead to efficient angular momentum transport and magnetic field amplification that could have important consequences on the dynamics of the postmerger remnant. The efficiency of the angular momentum transport is directly related to the timescale in which the black hole forms. Moreover, magnetic field amplification can generate large-scale structures that seem to favor jet formation and short gamma-ray bursts~\citep{Rezzolla:2011,Ruiz:2016,Combi:2023,Bamber:2024}. Another scenario where the MRI plays a role involves neutron star -- black hole mergers~\citep{Etienne:2012,Paschalidis:2015,Kiuchi:2015,Ruiz:2018,Christie:2019}. The instability sets in when an accretion disk is formed around the black hole after the merger. 

Due to the relevance of the MRI in many astrophysical settings, significant effort has been devoted to unraveling the physics of the instability and the resulting turbulent state. Seed perturbations can grow exponentially on timescales close to the rotational period. These perturbations take the form of so-called channel modes, which are pairs of vertically stacked layers in which the velocity and the magnetic field perturbations have radial and azimuthal components of (sinusoidally) alternating polarity. These modes have associated Maxwell and Reynolds stresses that lead to outward transport of angular momentum~\citep{Goodman:1994,Pessah:2006a,Pessah:2008}. 

The growth of the instability eventually terminates, resulting in the breakdown of the channels into small-scale turbulence.  The details of the processes involved in the saturation of the MRI, including the factor by which the seed perturbations are amplified, are not yet completely understood. Several authors~\citep{Hawley:1995,  Brandenburg:1995,Fleming:2000,Sano:2001,Sano:2004,Gardiner:2005,Pessah:2007,Vishniac:2009,Davis:2010,Murphy:2015,Rembiasz:2016a,Rembiasz:2016b,Hirai:2018,Gogi:2018} have provided further insight into the saturation of the MRI and the resulting nonlinear turbulent regime by performing numerical box simulations and also (semi-)global simulations of accretion disks~\citep{Sorathia:2010,Hawley:2011,Sorathia:2012}, fast rotating PNSs~\citep{Obergaulinger:2009,Mosta:2015,Reboul-Salze:2022} and BNS merger remnants~\citep{Kiuchi:2018,Shibata:2021}. 

\cite{Goodman:1994} presented a model for parasitic instabilities (PIs) that was further developed with local linear analyses by~\cite{Lesaffre:2009},~\cite{Latter:2009},~\cite{Pessah:2009}, and~\cite{Pessah:2010}, among others. This model provides a physical mechanism that explains the termination of the MRI and the onset of the nonlinear regime. Laminar channel flows can be unstable against PIs that can be of Kelvin-Helmholtz (KH) or tearing-mode (TM) type, depending on the value of kinematic viscosity and resistivity, that is,~nonideal effects. At the beginning of the exponential growth of the MRI, the effect of the PIs is negligible, since they grow much slower than the MRI. Nevertheless, the growth rate of the PIs is proportional to the amplitude of the MRI modes, which grows exponentially in time. This means that, at some point, the PIs  start growing much faster than the MRI modes, and they eventually disrupt the channel modes and saturate the MRI, leading to a turbulent regime. The predictions made by these analytical approaches have been tested by several numerical magnetohydrodynamic (MHD) simulations~\citep{Latter:2009,Latter:2010,Longaretti:2010,Lesur:2011,Murphy:2015,Rembiasz:2016a,Rembiasz:2016b,Hirai:2018}, but there are still some discrepancies between the analytical models and the numerical results. 

\cite{Pessah:2009} and~\cite{Pessah:2010} performed an analytical study in resistive-viscous MHD of the evolution of PIs by solving an eigenvalue problem with linear equations for these secondary instabilities. They exhaustively covered a huge parameter space to identify the fastest growing parasitic modes for different values of kinematic viscosity and resistivity. The authors \citep[and also][]{Latter:2009} made several assumptions to make the problem more tractable. The most notable simplifications are the consideration of the primary MRI mode as a time-independent background, and the assumption that the wavevectors of the parasitic modes are also time-independent (see Sect. 3.1 in~\cite{Pessah:2010} for a critical assessment of the assumptions involved).

In this work, we relax some simplifications made in previous studies to obtain a more accurate description of the evolution of PIs and a better estimate of the saturation of the MRI. Building on the approach in~\cite{Pessah:2010}, we derive a set of equations for the parasitic perturbations feeding of the fastest growing MRI mode for a fixed vertical magnetic field. However, here we account for the exponential growth of the MRI modes and the linear shear of the parasitic wavevector induced by differential rotation of the background flow\footnote{It is worth noting that already~\cite{Goodman:1994} presented as a case study the long-term dynamical evolution of a marginally unstable MRI mode including rotation and shear. However, they considered this primary mode as a static background for the parasitic modes.}.
By covering a dense parameter space, we identify the fastest secondary modes that lead to the saturation of the MRI. Using different values for the seed perturbations, we obtain amplification factors of the MRI that are similar to the ones obtained in the numerical simulations presented in~\cite{Rembiasz:2016b}.

The paper is organized as follows: in Sect.~\ref{sec::PI_eqs}, we state our working assumptions and present the equations for the PIs. 
In Sect.~\ref{sec::results}, we present the results of a systematic exploration to find the fastest parasitic modes responsible for the saturation of the MRI. We showcase the time evolution of several parasitic modes to provide physical intuition on the elements playing a role in the saturation process. In Sect.~\ref{sec::effective model}, we present our effective model for field amplification driven by the MRI and provide a expression for the MRI magnetic field at saturation. 
In Sect.~\ref{sec::comparison}, we compare our findings with numerical simulations. We discuss the implications of our findings in Sect.~\ref{sec::conclusions}.

\section{Parasitic instabilities feeding off exponential MRI modes}
\label{sec::PI_eqs}

To carry out a linear analysis of the parasitic modes feeding off the MRI channels, we need to treat those channels as part of the background fields, as in~\cite{Goodman:1994},~\cite{Pessah:2009} and~\cite{Pessah:2010}. This is a sensible approach during the MRI growth, since the amplitude of the channel modes is much larger than the parasitic ones. This implies that the primary instability (the MRI)  is not significantly affected by the secondary (parasitic modes) until they reach a similar amplitude. This approximation is bound to break down when the amplitudes involved are comparable. Nevertheless, here we consider that the primary MRI mode grows exponentially unimpeded. The novelty compared to previous studies is that we do consider that the parasites, with shearing time-dependent wavevectors, are feeding from a time-dependent MRI mode.

We consider the system of incompressible MHD equations governing the dynamics of the velocity $\boldsymbol{V}$ and magnetic $\boldsymbol{B}$ fields in the shearing box approximation 
\citep{Hawley:1995}
\begin{align}
    \partial_t  \boldsymbol{V} +(\boldsymbol{V}\cdot\nabla)\boldsymbol{V} & = -2 \boldsymbol{\Omega}\times \boldsymbol{V}+q\Omega^2\nabla (x^2)-\frac{1}{\rho}\nabla \left( P+\frac{B^2}{8\pi}\right) \\ 
    & + \frac{(\boldsymbol{B}\cdot\nabla)\boldsymbol{B}}{4\pi\rho}+\nu\nabla^2\boldsymbol{V}, \nonumber \\
    \partial_t \boldsymbol{B}+(\boldsymbol{V}\cdot\nabla)\boldsymbol{B} & = (\boldsymbol{B}\cdot\nabla)\boldsymbol{V}+\eta\nabla^2\boldsymbol{B}, \\
    \nabla \cdot \boldsymbol{V} & = 0, \label{v_div}\\ 
    \nabla \cdot \boldsymbol{B} & = 0\,.
\end{align}
Here, $\boldsymbol{\Omega} = \Omega\, \check{\bf z}$ stands for the angular frequency and $q$ is the shear parameter
\begin{equation}\label{shear}
    q \equiv -\frac{d\ln \Omega}{d \ln r}\Bigg \rvert_{r_0}\,,
\end{equation}
both of these quantities are evaluated at some fiducial radius $r_0$. 
$P$ stands for pressure, $\rho$ is the density, whereas $\nu$ and $\eta$ are the viscosity and resistivity. In what follows we simply write $\Omega$ for the local value of the angular frequency.

\subsection{Equations of motion for the parasitic modes}

The equations of motion for the PIs are obtained by seeking solutions for the total velocity and magnetic fields of the form~\citep{Goodman:1994}: 
\begin{equation}\label{state_vect_v2}  
  \left[\begin{array}{c}
         V_x  \\
        V_y \\
         V_z \\
        B_x \\
        B_y \\
        B_z 
        \end{array} \right]
        = \left[\begin{array}{c}
              0 \\
              -q\Omega x \\
              0 \\
              0 \\
              0 \\
              \bar{B}_z 
        \end{array}\right]
        + \left[\begin{array}{c}
              V^{\rm MRI}_x \\
              V^{\rm MRI}_y \\
              0 \\
              B^{\rm MRI}_x \\
              B^{\rm MRI}_y \\
              0 
        \end{array}\right]
        + \left[\begin{array}{c}
              v_x \\
              v_y \\
              v_z \\
              b_x \\
              b_y \\
              b_z
        \end{array}\right]\,.
\end{equation}
Here, we include, from left to right, the contribution of the background shear flow $\boldsymbol{V}_{\rm sh} = -q\Omega x \check{\bf y}$, the vertical magnetic field $\bar{B}_z$, the MRI fields (that also constitute the background dynamics of our problem), and the parasitic fields. For absent secondary perturbations, the vertical magnetic field $\bar{B}_z$ remains unchanged and the MRI grows exponentially unaffected. The former assumption holds in the incompressible limit because the MRI itself does not feed back into $\bar{B}_z$~\citep{Pessah:2006a}. This is a reasonable assumption in the context of local, shearing box simulations with an imposed vertical magnetic field, since the magnetic flux over the vertical boundaries is conserved when the usual azimuthally periodic boundary conditions are adopted. The second assumption is bound to break down when the parasitic mode amplitude is comparable to the amplitude of the MRI mode from which they feed. We address this issue in Sect.~\ref{sec::comparison}.

In the ideal, that is, for sufficiently small values of the viscosity $\nu$ and the resistivity $\eta$, incompressible MHD regime, the MRI evolves as an exact, nonlinear solution with a mode structure given by~\citep{Goodman:1994, Pessah:2006a, Pessah:2008}
\begin{align}\label{mri_fields}
    \boldsymbol{V}^{\rm MRI} & = V_0(t) \sin(K_{\rm MRI}z)[\cos(\theta_V) \check{\bf x}+\sin(\theta_V) \check{\bf y}],\\
    \boldsymbol{B}^{\rm MRI} & = B_0(t) \cos(K_{\rm MRI}z)[\cos(\theta_B) \check{\bf x}+\sin(\theta_B) \check{\bf y}]\,,
\end{align}
where $V_0(t) = V_0 e^{\gamma_{\rm MRI} t}$ and $B_0(t) = B_0 e^{\gamma_{\rm MRI} t}$ are the time-dependent MRI velocity and magnetic field amplitudes, respectively, $\check{\bf{x}}$ is the unit vector in the radial direction, $\check{\bf{y}}$ is the unit vector in the azimuthal direction, and $\theta_{\rm V}$ and $\theta_{\rm B}$ are the directions of the channels with respect to the radial direction. The fastest growing rate 
\begin{equation}\label{grate_mri}
    \gamma_{\rm MRI} = \frac{q}{2}\Omega\,,
\end{equation}
is attained for the MRI mode with vertical wavenumber 
\begin{equation}\label{fastest_k_general}
    K_{\rm MRI} = \sqrt{1 - \frac{(\kappa/\Omega)^4}{16}}\frac{\Omega}{\bar{v}^2_{{\rm A}z}}\,,
\end{equation}
where $\kappa \equiv \sqrt{2(2-q)}\Omega$ is the epicyclic frequency, and $\bar{v}_{{\rm A}z}$ corresponds to the Alfvén velocity. The physical structure of the fastest MRI mode is such that $\theta_{\rm V}=\pi/4$ and $\theta_{\rm B}=3\pi/4$.
The ratio between the MRI amplitudes in the ideal MHD limit, as found by~\cite{Pessah:2008}, is
\begin{equation}\label{ratio_ampl}
    \frac{V_0}{B_0/\sqrt{4\pi\rho}} = \sqrt{\frac{4-(\kappa/\Omega)^2}{4+(\kappa/\Omega)^2}}\,.
\end{equation}

As in~\cite{Pessah:2010}, we employ dimensionless variables defined in terms of the characteristic scales of length and time set by the background magnetic field and the local angular frequency: $L_0 \equiv \bar{v}^2_{{\rm A}z}/\Omega = \bar{B}_z^2/(4\pi\rho\Omega)$ and $T_0 \equiv 1/\Omega$. With this, $\bar{B}_z$ sets the scale for all magnetic and velocity fields. From now on, we omit the subscript ``MRI'' and use $K$, $\boldsymbol{V}$ and $\boldsymbol{B}$ to refer to the wavenumber, velocity, and magnetic fields associated with the fastest MRI mode. We absorb the factor $\sqrt{4\pi\rho}$ in the magnetic field, effectively working with Alfv\'en velocities.
 
The secondary, parasitic velocity and magnetic fields can be expressed as
\begin{align}
    \boldsymbol{v} & = \expo\boldsymbol{u}(t,z) \label{ansatz1_v}\,, \\
    \boldsymbol{b} & = \expo\boldsymbol{w}(t,z) \label{ansatz1_b} \,,
\end{align}
where the explicit temporal dependence of the horizontal wavevector is given by
\begin{equation}\label{kh}
    \kh = (k_x(0)+q\Omega k_yt)\check{\bf{x}}+k_y\check{\bf{y}}\,.
\end{equation}
This simply reflects the fact that wave crests are swept by the (linear) shear background flow, thereby increasing their wavenumber and rotating toward the radial direction, $\check{\bf{x}}$~\citep{Latter:2010,Mamatsashvili:2013}. 
 
We can exploit the incompressible nature of the flow and focus on the dynamics in the plane ($\boldsymbol{\check{k}_{\rm h}},\check{\boldsymbol{z}}$) \citep{Pessah:2010}. In fact, this condition restricts our problem to one single direction, since $i\kh\cdot\boldsymbol{v_{\rm h}} = -\partial_z v_z$.  We can furthermore eliminate the pressure by using the divergenceless condition for the velocity [Eq.~\eqref{v_div}].
Then, the evolution equations for the vertical components of the parasitic velocity and magnetic fields become
\begin{align} 
   (\partial_t & +i\kh\cdot\boldsymbol{V}+\nu\Delta)\Delta u_z-iK^2\kh\cdot\boldsymbol{V}u_z \label{v_z_nopress} \\ 
   + & q\sin2\theta\partial_z^2u_z- i\kh\cdot\boldsymbol{B}\Delta w_z+iK^2\kh\cdot\boldsymbol{B}w_z -\Delta \partial_z w_z = 0 \,, \nonumber \\
   (\partial_t & +i\kh\cdot\boldsymbol{V}+\eta\Delta)w_z-i\kh\cdot\boldsymbol{B}u_z-\partial_z u_z = 0\,, \label{b_z}
\end{align}
where we introduced the symbol $\Delta \equiv k_{\rm h}^2-\partial_z^2$, and $\theta$ is the time-dependent angle between the parasitic wavevector, $\kh$, and the radial direction in the counterclockwise sense. These equations are linear in the parasitic amplitudes in the incompressible regime.
The three-dimensional vector fields can be expressed solely in terms of their vertical components thanks to the divergenceless conditions, so that 
\begin{align}
    \boldsymbol{u} & = -\frac{\partial_z u_z}{ik_{\rm h}} \boldsymbol{\check{k}_{\rm h}}+u_z \check{\bf{z}} \,, \\
    \boldsymbol{w} & =   -\frac{\partial_z w_z}{ik_{\rm h}} \boldsymbol{\check{k}_{\rm h}}+w_z \check{\bf{z}} \,.
\end{align}

It is convenient to obtain the equations of motion for the vertical components of the parasitic modes in terms of Fourier series 
\begin{align}
    u_z & = \suma \alpha_n(t)\efou \,, \\ \label{u_fourier}
    w_z & = \suma \beta_n(t)\efou \,,
\end{align}
where $k_z$ is a parameter with $0 \leq k_z/K \leq 1/2$~\citep{Goodman:1994, Pessah:2010}. 
This leads to a band-diagonal system of differential equations for the temporal evolution of the Fourier coefficients
\begin{align}
    \partial_t \alpha_n(t) = & +i(n+k_z)K\beta_n(t) +\frac{q\sin2\theta}{2\Delta_n}                               (n+k_z)^2\alpha_n(t) \label{alpha_eq} \\ 
                             -&KV_0(t)\frac{\kh\cdot\check{\boldsymbol{V}}_0}{2\Delta_n}\big[\alpha_{n-1}(t)(\Delta_{n-1}-1)-\alpha_{n+1}(t)(\Delta_{n+1}-1)\big] \nonumber \\
                             +& iKB_0(t)\frac{\kh\cdot\check{\boldsymbol{B}}_0}{2\Delta_n}\big[\beta_{n-1}(t)(\Delta_{n-1}-1)+\beta_{n+1}(t)(\Delta_{n+1}-1)\big]  \nonumber \\
                             -&\nu K^2\Delta_n\alpha_n(t) \nonumber \,, \\
    \partial_t \beta_n(t) = & -KV_0(t)\frac{\kh\cdot\check{\boldsymbol{V}}_0}               {2}\big[\beta_{n-1}(t)-\beta_{n+1}(t) \big] \label{beta_eq} \\
                             +&iKB_0(t)\frac{\kh\cdot\check{\boldsymbol{B}}_0}{2}\big[\alpha_{n+1}(t)+\alpha_{n-1}(t) \big] + i(n+k_z)K\alpha_n(t) \nonumber \\
                             -&\eta K^2\Delta_n\beta_n(t) \nonumber \,,
\end{align}
 where $\check{\boldsymbol{V}}_0 \equiv \cos(\theta_V) \check{\bf x}+\sin(\theta_V) \check{\bf y}$, $\check{\boldsymbol{B}}_0 \equiv \cos(\theta_B) \check{\bf x}+\sin(\theta_B) \check{\bf y}$ and $\Delta_n \equiv k_{\rm h}^2+(n+k_z)^2$.   

In writing the previous equations, we have scaled the time-dependent Fourier coefficients in terms of the initial amplitude of the MRI channel, $B_0$ ($\alpha_n \rightarrow \alpha_n/B_0$, $\beta_n \rightarrow \beta_n/B_0$), and the wavenumbers $\kh$ and $k_z$ in terms of the MRI one, $K$ ($\kh \rightarrow \kh/K$, $k_z \rightarrow k_z/K$). We have also used the Euler expressions $\sin(Kz) = (e^{iKz}-e^{-iKz})/2i$ and $\cos(Kz) = (e^{iKz}+e^{-iKz})/2$.

\subsection{The initial value problem}
\label{subsec::initial_data}

To integrate the set of differential equations, we must provide appropriate initial values. The initial Fourier amplitudes of the PIs,  that is,~the coefficients $\alpha_n(0)$ and $\beta_n(0)$, can be obtained using the equations from~\cite{Pessah:2010}, which define a linear, eigenvalue problem where the eigenvalues correspond to the growth rate of the PIs and the eigenvectors' components are the values of $\alpha_n$, $\beta_n$. These equations are:
\begin{align}
    \frac{s}{KB_0}\alpha_n(0)  = & +\frac{i}{B_0}(n+k_z)\beta_n(0) \label{alpha0_eq} \\
                            -&\frac{\kh\cdot\check{\boldsymbol{V}}_0}{2\Delta_n}  \frac{V_0}{B_0}\big[\alpha_{n-1}(0)(\Delta_{n-1}-1)-\alpha_{n+1}(0)(\Delta_{n+1}-1)\big] \nonumber \\
                             +&i\frac{\kh\cdot\check{\boldsymbol{B}}_0}{2\Delta_n}\big[\beta_{n-1}(0)(\Delta_{n-1}-1)+\beta_{n+1}(0)(\Delta_{n+1}-1)\big] \nonumber \\
                             -&\frac{\nu K^2}{KB_0}\Delta_n \alpha_n(0)\nonumber \,, \\
    \frac{s}{KB_0}\beta_n(0) = & -\frac{\kh\cdot\check{\boldsymbol{V}}_0}{2}\frac{V_0}{B_0}\big[\beta_{n-1}(0)-\beta_{n+1}(0)\big]\label{beta0_eq} \\
                             +&i\frac{\kh\cdot\check{\boldsymbol{B}}_0}{2}\big[\alpha_{n-1}(0)+\alpha_{n+1}(0)\big]+i\frac{1}{B_0}(n+k_z)\alpha_n(0) \nonumber \\
                             -&\frac{\eta K^2}{K B_0}\Delta_n\beta_n(0)  \nonumber \,,
\end{align}
where $s$ is the growth rate of the parasitic modes. In the ideal MHD limit, the last term in the right-hand-side of both equations vanishes. 

This set of equations corresponds to the problem solved in~\cite{Pessah:2010} under the assumption that the MRI primary mode and the parasitic wavenumber $\kh$ are both time-independent. These equations allow us to obtain an approximate initial solution of a given parasitic mode. Considering a value for the initial MRI amplitude $B_0$ and a fixed parasitic wavenumber $\kh(t=0)$, we obtain the parasitic eigenvector corresponding to the eigenvalue with the largest real contribution (fastest PI mode). This provides a set of $\alpha_n(0)$ and $\beta_n(0)$ values that can be used as initial values to solve the time-dependent Eqs.~\eqref{alpha_eq} and~\eqref{beta_eq}, where the amplitude of the MRI $B_0(t)$ and the parasitic wavenumber $\kh(t)$ are time-dependent\footnote{One possible option is to employ white noise,  namely, equal power in each Fourier mode, to seed the secondary modes. We choose to excite the secondary eigenmodes because this gives us good control over their onset to search for the fastest growing secondaries systematically. Exciting a broad spectrum of perturbations at once would only mask the fastest growing parasite initially.}.

Since the parasitic equations are linear, we must provide the initial amplitude for the PI modes, $v_0 \equiv v(t=0)$. We choose to parameterize the initial amplitude of the parasitic mode with the velocity $v_0$ because in ideal MHD these instabilities correspond to KH modes\footnote{The reason to parameterize the amplitude of the primary MRI mode by its magnetic field $B_0$ is that the ratio $V_0/B_0$ may become vanishingly small in limiting nonideal regimes.}. Realize that the value of the parasitic magnetic field $b_0$ is determined by the PI eigenvalue problem. 

To summarize, under our working assumptions, the dynamics of a parasitic mode with an initial wavenumber $\kh(0)$ feeding off the fastest, exponentially growing MRI mode depends exclusively on the initial amplitude of the MRI mode, $B_0$, and the initial amplitude of the parasitic mode, $v_0$. The background magnetic field that determines the dynamics of the fastest growing MRI mode provides an overall scale.

\subsection{Solution for the parasitic instabilities}

Having solved for the temporal evolution of the Fourier coefficients (see Appendix~\ref{app:time_ev}), the evolution of the parasitic velocity and magnetic fields in physical space is obtained as
\begin{align}
    v_z(t;\boldsymbol{x}) & = \suma \alpha_n(t) \efouv\expo \,, \\
    b_z(t;\boldsymbol{x}) & = \suma \beta_n(t) \efouv \expo \,.
\end{align}

Since the horizontal component in the direction of $\kh$ is proportional to the vertical component thanks to the divergence-free condition, the velocity and magnetic field parallel to the plane defined by ($\boldsymbol{\check{k}_{\rm h}},\check{\boldsymbol{z}}$) are given by
\begin{align}
   \boldsymbol{v}(t;\boldsymbol{x}) & = -\frac{1}{k_{\rm h}}\suma (n+k_z)\alpha_n(t) \efouv\expo \boldsymbol{\check{k}_{\rm h}} \label{v_fou}\\
     & + \suma \alpha_n(t) \efouv\expo\check{\bf{z}}\,, \nonumber \\
    \boldsymbol{b}(t;\boldsymbol{x}) & = -\frac{1}{k_{\rm h}}\suma (n+k_z)\beta_n(t) \efouv\expo \boldsymbol{\check{k}_{\rm h}} \label{b_fou}\\
    & +\suma \beta_n(t) \efouv \expo \check{\bf{z}}\,. \nonumber
\end{align}

These expressions are in essence the same as in \cite{Pessah:2010}, except for the crucial difference that the Fourier coefficients depend explicitly on time taking into consideration the exponential growth of the MRI and the linear shear of the parasitic wavevector, $\kh(t)$, by the local background flow. In Appendix~\ref{app:phys_structure}, we depict the physical structure of both the parasitic velocity and magnetic fields, and also including the primary MRI fields.  

\subsection{Saturation of the MRI based on parasitic mode amplitudes}

As long as the amplitude of the PIs remains small one can justify neglecting their effect on the MRI mode from which they feed. This approximation eventually starts to break down and the MRI saturates.  An estimate for this saturation amplitude can be obtained by seeking the time $t_{\rm sat}$ that it takes to the fastest parasite to reach a kinetic energy comparable to the fastest MRI mode\footnote{This is motivated by the fact that in ideal MHD the parasitic modes that grow fastest are of the KH type. For sufficiently large resistivity, where TMs become important, it may be appropriate to consider the magnetic energy of the modes involved.}. In practice, we assume that the MRI mode (and also the PI mode) saturates when the average velocity of the fastest parasite reaches a certain fraction $\epsilon \sim 1$ of the MRI velocity~\citep{Latter:2010,Rembiasz:2016a}:
\begin{equation}\label{thresh_sat}
    v(t_{\rm sat}) = \epsilon V(t_{\rm sat})\,, 
\end{equation}
where the MRI primary mode grows according to 
\begin{equation}\label{energy_MRI}
    V(t) \equiv |\bar{\boldsymbol{V}}(t)| = \frac{V_0(t)}{\sqrt{2}}\,, 
\end{equation}
whereas the average velocity of the PI modes can be computed from the Fourier amplitudes $\alpha_n(t)$:
\begin{equation}\label{v_PI}
    v(t) \equiv |\bar{\boldsymbol{v}}(t)| =  B_0  \sqrt{\suma \Bigg[1+\frac{(n+k_z)^2}{k_{\rm h}^2}\Bigg]|\alpha_n(t)|^2}\,.
\end{equation}

Before presenting the results of a systematic study of the saturation amplitude for the MRI in Sect.~\ref{sec::results}, it is useful to introduce the following two quantities. We define the instantaneous growth rate of a parasitic mode as
\begin{equation}\label{grate_pi}
    \gamma_{\rm PI}(t) = \frac{1}{\Delta t}\frac{v(t)-v(t-\Delta t)}{v(t-\Delta t)}\,,
\end{equation}
using a time interval $\Delta t$. We also define the amplification factor $\mathcal{A}$ in terms of the ratio between the volume-averaged Maxwell stress tensor and the initial vertical magnetic field~\citep{Rembiasz:2016b}: 
\begin{equation}\label{ampl_fact_general}
    \mathcal{A} \equiv \sqrt{\bar{\mathcal{M}}_{xy}^{\rm sat}} = \frac{B_0(t_{\rm sat})}{\sqrt{2}\bar{B}_z}\sqrt{|\cos\theta_{\rm B}\sin\theta_{\rm B}|}\,.
\end{equation}
In ideal MHD this corresponds to 
\begin{equation}\label{ampl_fact}
    \mathcal{A} = \frac{1}{2}\frac{B_0(t_{\rm sat})}{\bar{B}_z}\,.
\end{equation}

\section{Results}
\label{sec::results}

\subsection{Searching for the fastest parasitic modes}

We must first characterize the fastest growing MRI mode off which the parasites will feed.
Since numerical simulations are not completely dissipation-less, we consider very small values of the dissipation coefficients. For all practical purposes, we are otherwise working very close to the ideal MHD regime. We therefore set the (dimensionless) viscosity to $\nu = 10^{-3}$ and the resistivity to $\eta = 10^{-2}$. The fastest MRI mode (for Keplerian shear with $q = 1.5$) is characterized by a vertical wavenumber $K \approx 0.96$ and angles $\theta_{\rm V} = 44.5^{\circ} \approx 45^{\circ}$ and $\theta_{\rm B} = 134.7^{\circ} \approx 135^{\circ}$, as shown in~\cite{Pessah:2010}.  In addition, the ratio between the MRI amplitudes is $V_0/B_0 \approx 0.77$. 

To obtain a systematic understanding of the saturation of the MRI we need to sweep parameter space and identify the parasite that reaches a comparable amplitude fastest. To accomplish this, we solve a large number of initial value problems as described above. We consider a range of initial values for the PI wavevectors $\kh(0) = k_x(0)\check{\bf{x}}+ k_y\check{\bf{y}}$ such that\footnote{We focus on parasitic modes with the same vertical periodicity than the MRI modes, that is, $k_z = 0$, since are expected to grow faster~\citep{Goodman:1994,Pessah:2010,Rembiasz:2016a,Hirai:2018}.}
\begin{eqnarray}\label{kh_ini}
    k_x(0) & = & \{-1,-1.125,-1.25,...,-15\},  \\
    k_y & = & \{0.1,0.1125,0.125,...,0.7\} \,.
\end{eqnarray}

We also tested the evolution with initial values of $k_x(0)$ and $k_y$ 
from other regions of the parameter space, namely,~positive $k_x(0)$ and smaller $k_y$. However, the primary MRI mode saturated very late, or did not even saturate. Since the parasitic growth rate is proportional to the MRI amplitude (see Sect.~\ref{subsec:anal_formula}), the parasitic modes need enough time before being able to grow super-exponentially. This entails allowing the primary MRI mode to reach a sufficiently large amplitude.
If the initial value of $k_x$ is positive, the mode will swing through $\theta_{\rm V}$ too early (see Fig.~\ref{fig:theta_cone}). Moreover, a large enough value of $k_y$ is needed, of the order $\mathcal{O}(0.1)$, as shown by~\cite{Pessah:2009},~\cite{Pessah:2010}.
We must also provide the initial amplitudes for the fastest MRI mode $B_0$ and the initial parasitic perturbation $v_0$. For this we consider
\begin{eqnarray}\label{ini_ampl}
    B_0 & = & \{1,5,10,50,100\} \times 10^{-4},  \\
    v_0 & = & \{0.1,0.5,1,5,10\} \times 10^{-4}\,.
\end{eqnarray}

\begin{figure*}[h]
\centering
\includegraphics[width=17cm]{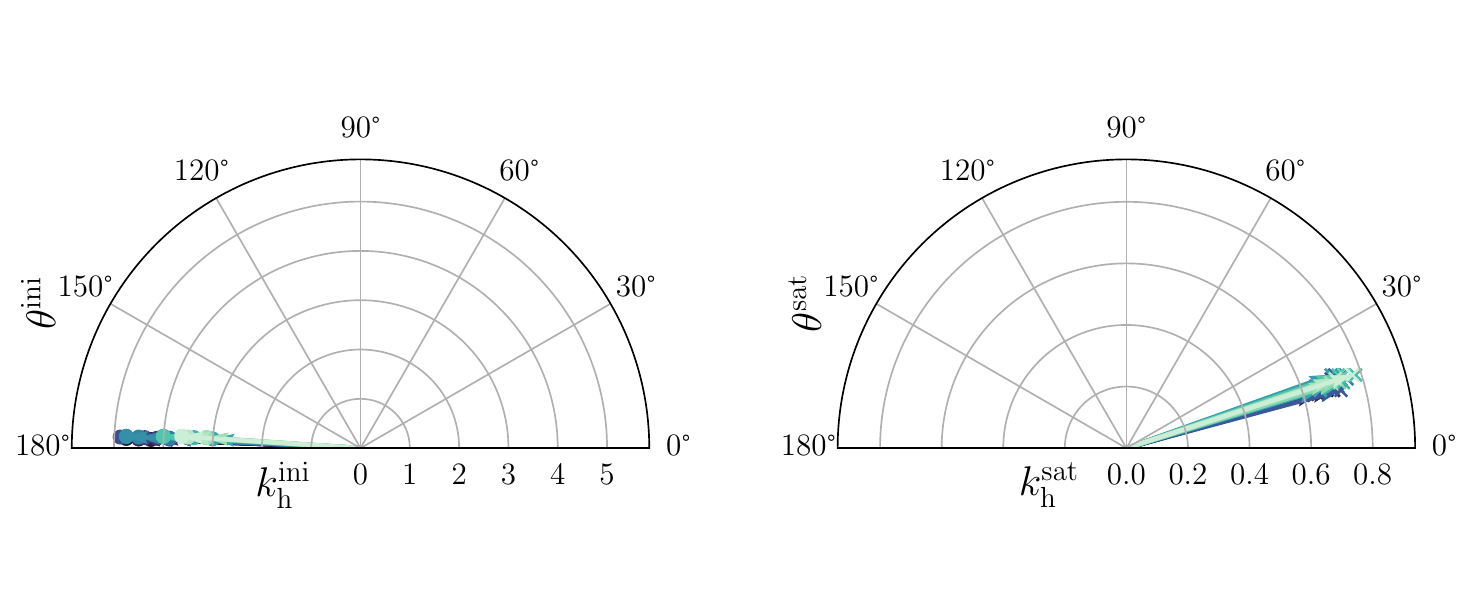}
\caption{Representation of the parasitic wavevectors $k_{\rm h}$ at $t=0$ and at saturation, $t= t_{\rm sat}$, for all the runs in Table~\ref{tab:runs}. The initial modulus $k_{\rm h}^{\rm ini}$ ranges between 3 and 5, but the initial angle $\theta^{\rm ini}$ is very close to $180^{\circ}$ in all cases. The wavevectors at saturation have almost the same angle $\theta^{\rm sat}$ at around $18^{\circ}$, and the modulus $k_{\rm h}^{\rm sat}$ lies around 0.7.}
\label{fig:kh_all_b0}
\end{figure*}

The properties of the fastest growing parasitic modes and MRI saturation amplitudes are summarized in Table~\ref{tab:runs} in Appendix~\ref{app:runs}. The analysis of these results leads to several interesting conclusions. The fastest growing modes possess initial horizontal wavevectors $\sim 3-5$ times larger than $K$, with $k_x(0) \gg k_y$, and values at saturation at around $0.6-0.7$, with $\theta^{\rm sat} \lesssim 20^{\circ}$. Fig.~\ref{fig:kh_all_b0} shows the initial and final wavevectors of the fastest PI modes in each of these runs. The amplification factors take values within the range $\sim 35 - 55$, depending on the initial parasitic amplitude. Although not shown here, many parasitic modes grow almost as fast as the fastest ones, leading to similar amplification factors. Even though the initial PI amplitudes are larger than the primary modes in some cases, the results also seem valid since the early growth of the PIs is much slower than that of the MRI. 

\begin{figure}[h]
\centering
\resizebox{\hsize}{!}{\includegraphics{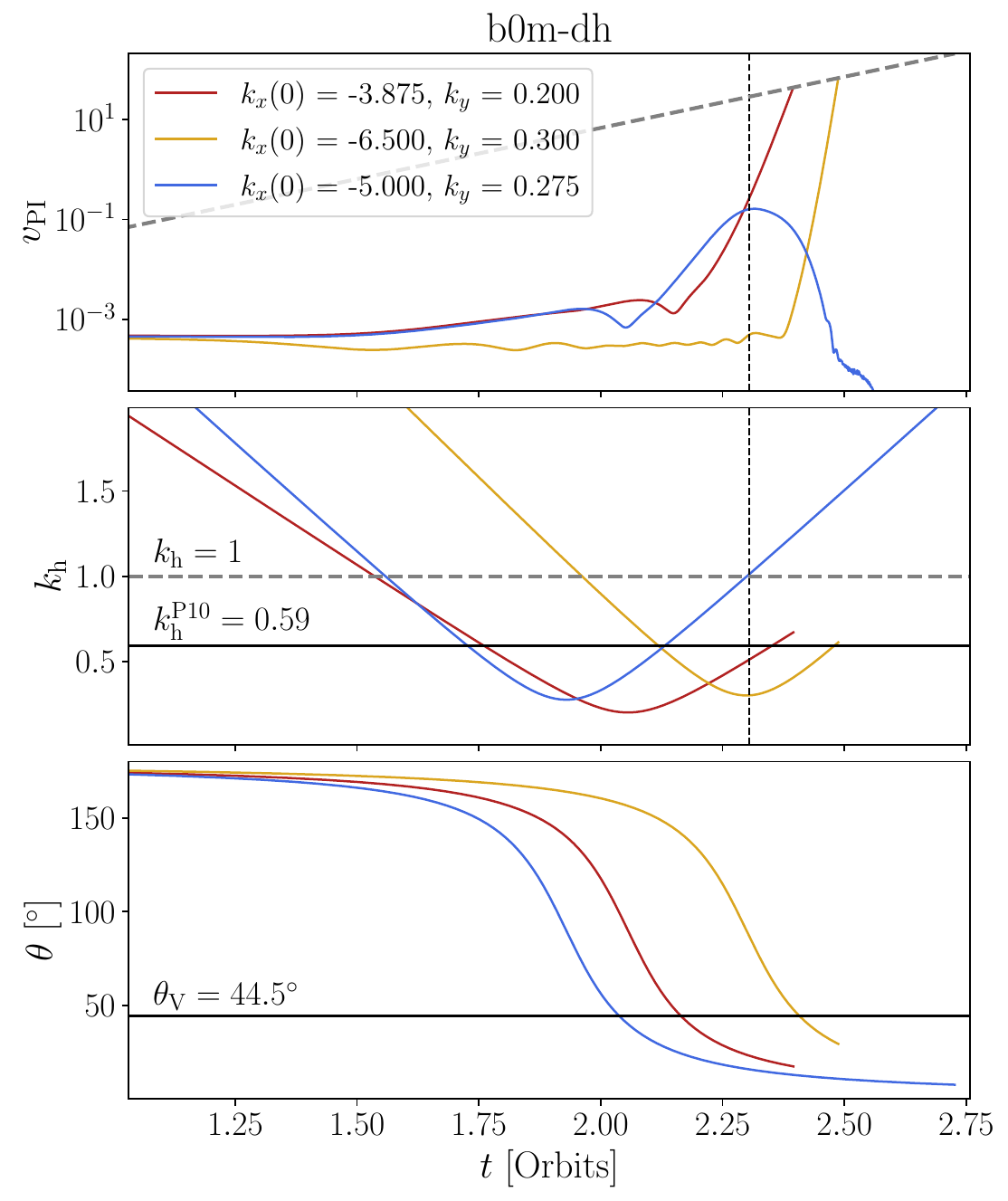}}
\caption{Top panel: time evolution of the averaged velocity, $v_{\rm PI}$, of three parasitic modes from the run \texttt{b0m-dh}: the fastest (red), another that saturates at later times (yellow), and one mode that does not saturate (blue). Middle panel: time evolution of the parasitic wavevector $k_{\rm h}$. Bottom panel: time evolution of the angle $\theta$ between the wavevector $\kh$ and the radial direction, $\check{\bf{x}}$. The modes start growing faster when $k_{\rm h}$ and $\theta$ approach the values found in~\cite{Pessah:2010}, denoted as $k_{\rm h}^{\rm P10}$ and $\theta^{\rm P10}$, respectively.}
\label{fig:PI-ev_theta_kh}
\end{figure}

\begin{figure}[h]
\centering
\resizebox{\hsize}{!}{\includegraphics{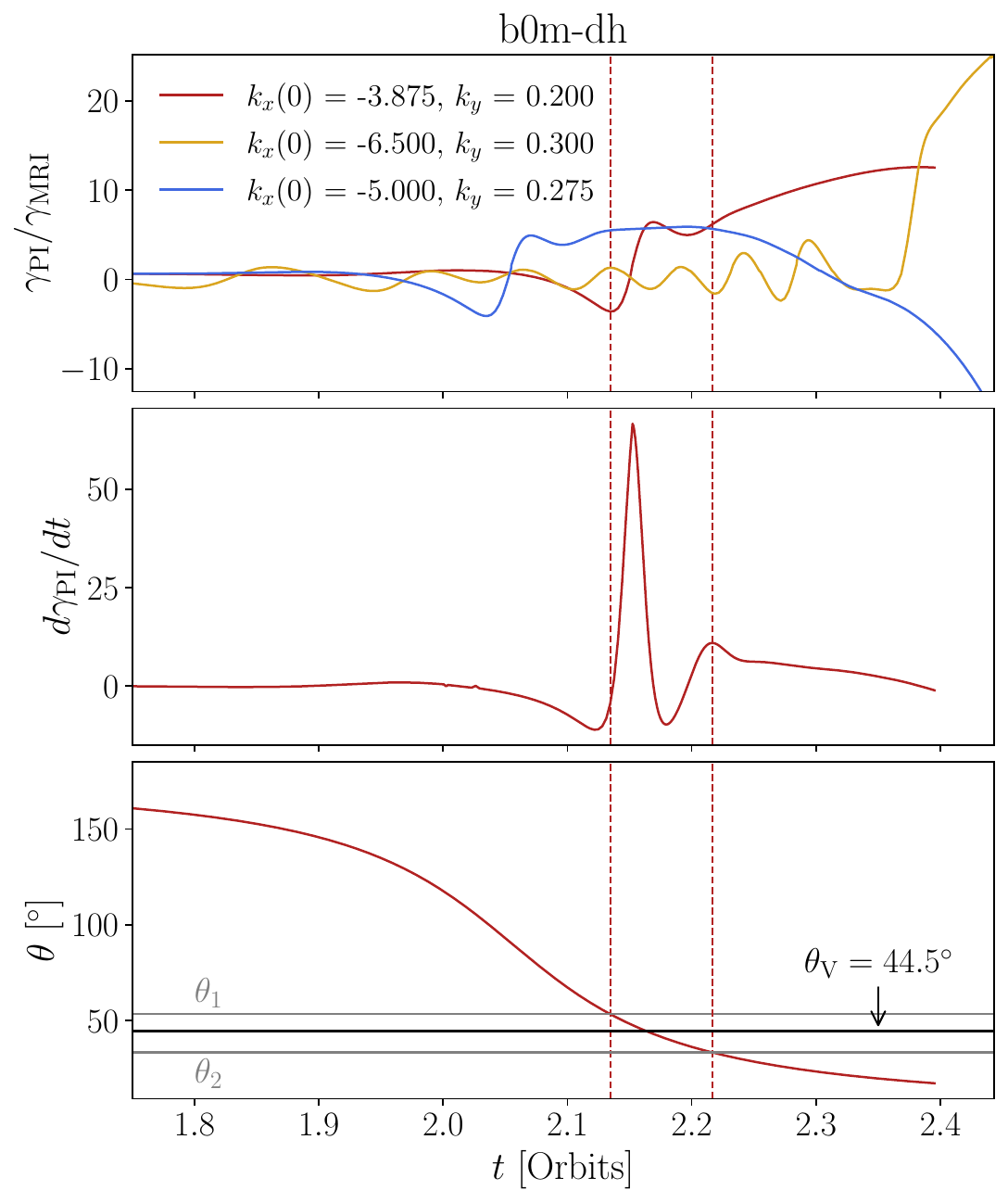}}
\caption{Top panel: time evolution of the normalized growth rate for the same modes as in Fig.~\ref{fig:PI-ev_theta_kh}. The growth rate stays considerably low up to a certain point, where it starts increasing to become more than ten times larger than the growth rate of the MRI (for the cases that reach saturation). Middle panel: evolution of the time derivative of $\gamma_{\rm PI}$ for the fastest mode. We draw vertical lines where the derivative takes positive values and at its last local maximum before saturation. Between these times, the parasitic mode grows the fastest. In the bottom panel, we show the evolution of the wavevector angle $\theta$. We depict horizontal gray lines showing $\theta_1$ and $\theta_2$, i.e., the values of $\theta$ that indicate the fastest growth of the parasitic mode.}
\label{fig:grates}
\end{figure} 

\subsection{Understanding the results}

To shed light on the dynamics characterizing the behavior of the fastest growing parasites we examine in detail the evolution of some of the modes from run \texttt{b0m-dh} in Figs.~\ref{fig:PI-ev_theta_kh},~\ref{fig:grates}. We focus our attention on the mode that is first to reach the same amplitude as the MRI, that is to say, the fastest parasitic mode (red line); a mode that takes a bit longer to accomplish the same (yellow line), and a mode that grows initially but then stalls and later decays (blue line).   

The top panel of Fig.~\ref{fig:PI-ev_theta_kh} depicts the time evolution of the velocity computed from Eq.~\eqref{v_PI} for these three modes. 
The vertical velocity shear flow induced by the MRI is maximum in the direction given by  $\theta_{\rm V}$. Examining the temporal evolution of $k_{\rm h}$ and $\theta$ is key to understanding why certain modes grow faster than others.  The parasitic modes that can most effectively tap into this energy source, are those that, as their wavevectors are swept by the background flow according to Eq.~\eqref{kh}, reach the direction $\theta_{\rm V}$ with an optimal wavenumber. The velocity of the mode that does not saturate starts to decrease close to the time when its wavenumber $k_{\rm h}$ becomes larger than unity (middle panel). This seems in agreement with the results known from the time-independent calculations in~\cite{Goodman:1994} and~\cite{Pessah:2010}, that show that parasitic modes with $k_{\rm h} >1$ do not grow.  For the  fastest growing mode (in red), $\theta$ starts decreasing before the slower mode (in yellow). {This allows the fastest mode to start growing before the mode in yellow, when the MRI velocity is $\sim 10^3-10^4$ times larger than the parasitic velocity, reaching values of $\theta \approx \theta^{\rm P10} = \theta_{\rm V}$ and $k_{\rm h} \approx k_{\rm h}^{\rm P10} = 0.59$~\citep{Pessah:2010} at earlier times. There is also a transient amplification of the parasitic velocity before the super-exponential growth that occurs when $\theta \approx \theta_{\rm V}$. This happens as the parasitic wavevector swings through from leading to trailing. 

We showcase in Fig.~\ref{fig:grates} the time evolution of the growth rate of these PI modes (computed with Eq.~\eqref{grate_pi}), normalized by the growth rate of the MRI from Eq.~\eqref{grate_mri}. The growth rate of the fastest mode starts increasing monotonically before the other mode that also saturates. Some modes reach a larger growth rate, but they saturate later because they get excited also later. The growth rate of the fastest growing mode starts increasing fast above 0 at $t \approx 2.15$ orbits and then it continues growing at a slower rate. The middle panel shows the evolution of its time derivative. The region where the derivative takes positive values and increases faster coincides with values of $\theta$ (bottom panel) around $\theta_{\rm V} = 44.5^{\circ}$. The rapid decrease of the derivative of the growth rate before the period of fast growth is due to the stabilizing effect of the MRI magnetic field, which is perpendicular to the MRI velocity field\footnote{We have tested this is the case by excluding the MRI magnetic field from the equations for the parasitic modes and confirming that the associated parasitic modes grow monotonically.}. This behavior is also observed in the other cases from Table~\ref{tab:runs}, and also for other modes that saturate at later times (see also Fig.~\ref{fig:effective_grate}). 

\begin{figure}[h]
\centering
\resizebox{\hsize}{!}{\includegraphics{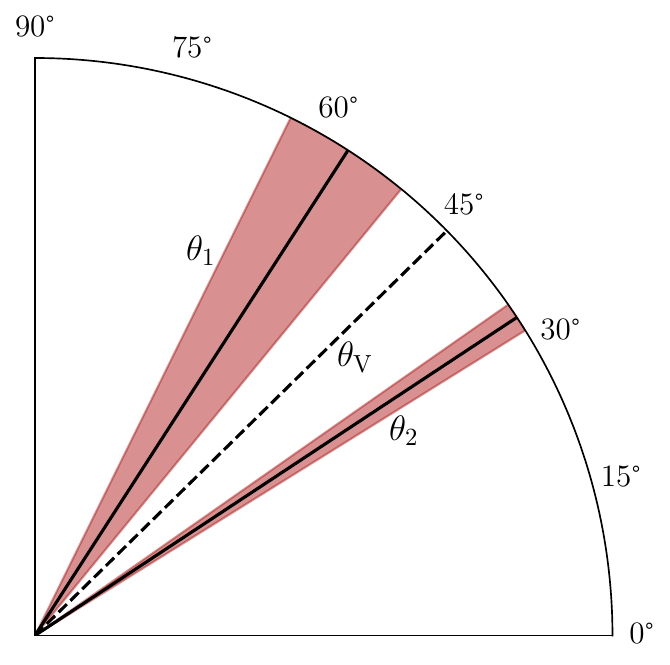}}
\caption{Values of the angle $\theta$ for which the PIs grow super-exponentially. The rapid increase of the secondary modes occurs between $\theta_1$ and $\theta_2$, depicted with solid black lines. These lines correspond to the mean value of these angles for the runs from Table~\ref{tab:runs}. The red-shaded regions represent the 1-$\sigma$ deviation. The black dashed line stands for the direction of the MRI velocity field, $\theta_{\rm V} = 44.5^{\circ}$.}
\label{fig:theta_cone}
\end{figure} 

Figure~\ref{fig:theta_cone} shows that the rapid increase of the parasitic modes coincides with the time-dependent wavevector $\kh$ being aligned with the MRI velocity field. When the growth rate starts increasing monotonically at $\theta_1$ (cf.~Fig.~\ref{fig:grates}), the direction of $\kh$ is getting close to $\theta_{\rm V}$. The mode grows faster and its increase slows down when $\theta$ falls below $\theta_{\rm V}$. The angle $\theta_2$ of Fig.~\ref{fig:theta_cone} corresponds to the last turning point of the growth rate before saturation. The values of $\theta_1$ and $\theta_2$ are given by the vertical dashed lines of Fig.~\ref{fig:grates} (see also the horizontal gray lines in the bottom panel). The black solid lines for $\theta_1$ and $\theta_2$ correspond to the mean value of these angles from the runs of Table~\ref{tab:runs}, and the shaded regions in red depict the 1-$\sigma$ deviation. As seen in Fig.~\ref{fig:theta_cone}, the values of $\theta_2$ are almost the same for all runs, which means that, independently from the initial amplitude of the instabilities, the fastest parasitic modes start increasing at a reduced rate when $\theta \gtrsim 30^{\circ}$. Alternatively, there seems to be a larger dispersion for $\theta_1$, with an average value of $\overline{\theta}_1 \approx 60^{\circ}$. Thus, the modes start their phase of super-exponential growth at an  angle $\theta = \theta_1 > \theta_{\rm V}$. After the braking in the parasitic growth, the fastest mode eventually saturates. This result is consistent with the findings made by~\cite{Pessah:2010}, even though the saturation criterion is different. In~\cite{Pessah:2010}, the saturation occurs when the parasitic growth rate equals the MRI one, and our criterion is the one given in Eq.~\eqref{thresh_sat}. When the parasitic amplitude reaches a similar value than the MRI channel's, the parasitic growth rate is already several times larger than the MRI growth rate (see Fig.~\ref{fig:grates}). Furthermore, with the current approach, it takes more time to the MRI to saturate, leading to a smaller $\theta^{\rm sat}$ and larger $k_{\rm h}^{\rm sat}$. This is in agreement with the predictions from \cite{Latter:2010}, who stated that it should take more time to saturate due to the inclusion of the background shear in the equations and the time dependence of $\kh$. 

\section{An effective model for the amplification factor}
\label{sec::effective model}

\begin{figure*}[t]
\centering
\includegraphics[width= 17cm]{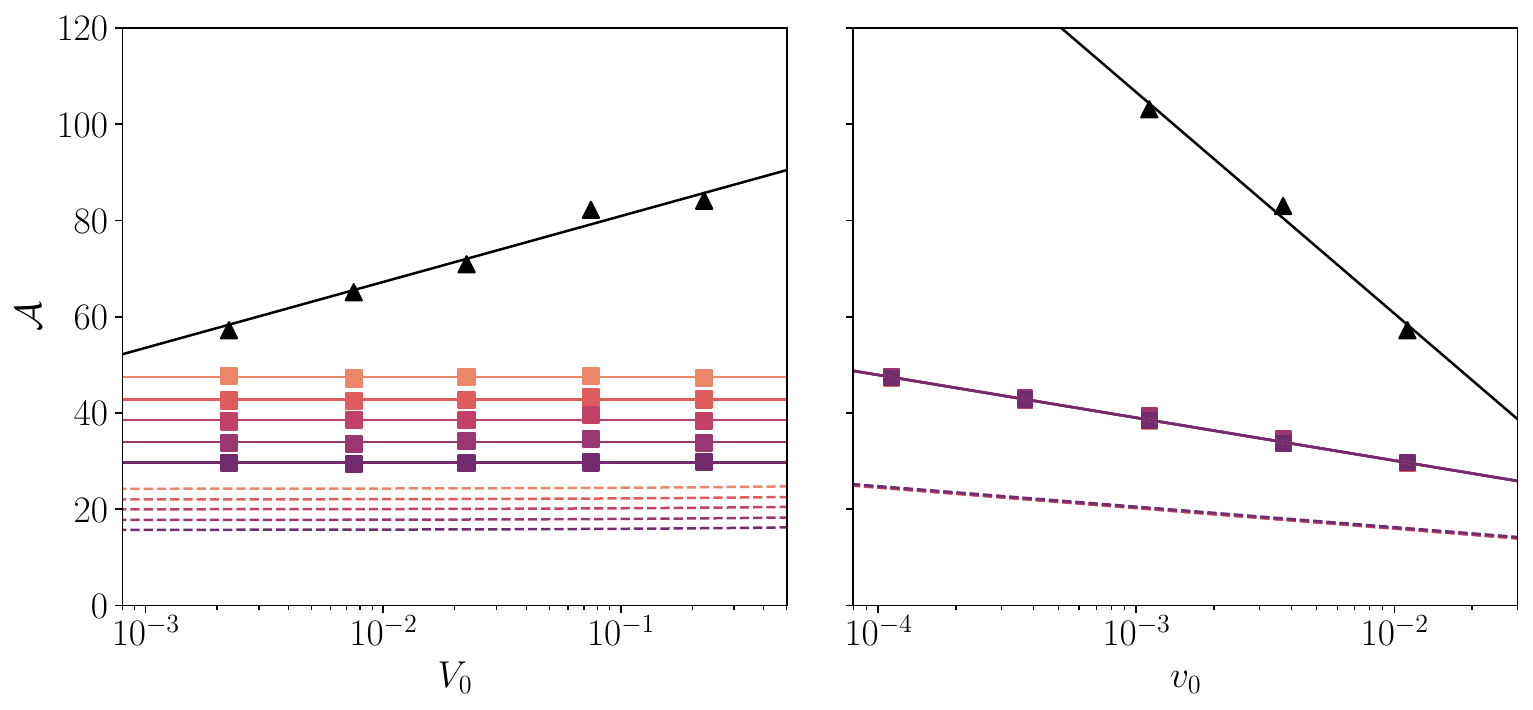}
\caption{Left panel: dependence of the amplification factor, defined in Eq.~\eqref{ampl_fact}, on the amplitude of the initial MRI velocity field, $V_0$. Right panel: dependence of the amplification factor on the initial PI velocity field, $v_0$. We use the runs from Table~\ref{tab:runs2} (solid lines with squares). Each color represents a different initial PI amplitude, $v_0$, where the darker colors stand for larger values of $v_0$. The solid lines represent the fit from Eq.~\eqref{ampl_coeffs}. The dashed lines stand for the amplification factor computed with Eq.~\eqref{ampl_approx}, i.e., assuming no background shear. The black triangles correspond to the results from the numerical simulations of~\cite{Rembiasz:2016b}, and the black solid line is the linear fit from Eq.~\eqref{fit_sims}.}
\label{fig:ampl_1}
\end{figure*}

Appealing to physical intuition and guided by our numerical results, we here seek to provide a simple expression for the amplification factor obtained when the fastest parasite, being advected by the background shear flow, is able to match the amplitude of the fastest, exponentially growing MRI mode.

In this Section, we employ a different value of the shear parameter, $q = 1.25$, instead of the Keplerian value $(q = 1.5)$, in order to make a better comparison with the results from the numerical simulations of~\cite{Rembiasz:2016b}, who use $q=1.25$. We also consider the fastest MRI mode associated to this value of $q$. This mode is characterized by a vertical wavenumber $K \approx 0.93$, angles $\theta_{\rm V} = 45^{\circ}$ and $\theta_{\rm B} = 135^{\circ}$, and a ratio of MRI aplitudes $V_0/B_0 \approx 0.67$. All these values are very similar to those for $q = 1.5$. 

The analysis of the results presented in Table~\ref{tab:runs2} (employing $q=1.25$) and illustrated in Fig.~\ref{fig:ampl_1} reveals that the amplification factor is insensitive to the initial amplitude of the MRI mode while it may change by a factor of a few when the initial amplitude of the parasitic modes varies by a few orders of magnitude. The results from Table~\ref{tab:runs} (using $q = 1.5$) are not depicted in Fig.~\ref{fig:ampl_1}, but are almost identical, meaning that the slight change in the shear parameter $q$ does not have notable effects.

We can make use of this result to find an approximate expression for the amplification factor which is physically motivated. We proceed in two steps. First, we find the expression for the amplification factor ignoring the fact that the parasitic wavevector $\kh$ is advected by the background shear. We account for this effect in a second step as described below. 

\subsection{Incorporating the exponential growth of the MRI}
\label{subsec:anal_formula}

Building on the results presented in \cite{Pessah:2010},  \cite{Rembiasz:2016b} derived an approximate expression for the amplification factor assuming that the parasitic growth rate is given by
\begin{equation}\label{g_pi_pessah2010}
    \gamma_{\rm PI}^{\rm P10} = \sigma K V_0(t)\,,
\end{equation}
where $\sigma = 0.27$~\citep{Pessah:2010} and considering that $\gamma_{\rm PI}^{\rm P10} \equiv \dot{v}(t)/v(t)$. This leads to an analytical expression for the velocity of the parasitic mode
\begin{equation}\label{v_pi_p10}
    v(t) = v_0\exp \left[\frac{\sigma K V_0}{\gamma_{\rm MRI}}\left(e^{\gamma_{\rm MRI}t}-1\right)\right]\,.
\end{equation}

Equating $v(t_{\rm sat}) =\epsilon V(t_{\rm sat})$, and using the parasitic velocity from Eq.~\eqref{v_pi_p10}, leads an analytical expression for the amplification factor \citep{Rembiasz:2016b}
\begin{equation}\label{ampl_p10}
    \mathcal{A}-\frac{1}{2\sigma}\ln\mathcal{A} = \frac{1}{2\sigma}\left[ - \ln(v_0) +\ln\left(\epsilon \sqrt{\frac{4q}{4-q}}\right) \right]+\sqrt{\frac{4-q}{4q}}V_0\,.
\end{equation}
This equation can be solved using the Lambert$W$ function~\citep{Corless:1996}, which has known asymptotic approximations~\citep[see, e.g.,][for an application]{Latter:2016}. However, we prefer approximating it to a simple analytical formula.
It can be seen that the amplification factor scales with the logarithm of the initial parasitic amplitude and also linearly with the initial MRI amplitude. Since initially $V_0 \ll 1$, the amplification factor should be almost independent of the initial MRI channel amplitude. This behavior is observed in Fig.~\ref{fig:ampl_1} (left panel), and the linear dependence with the logarithm of the parasitic amplitude can be seen in the right panel. The logarithmic dependence in the parameters $q$ and $\epsilon$ implies that these values do not strongly affect the amplification factor. Similarly, one could obtain an expression for the saturation time, $t_{\rm sat}$, which can provide another interesting diagnostic.

On the left-hand side of Eq.~\eqref{ampl_p10} there is a term involving $\ln\mathcal{A}$ that prevents us from finding a close solution for the amplification factor. However, as shown in Fig.~\ref{fig:ampl_1}, the spread in $\mathcal{A}$ is not big, which means that $\ln\mathcal{A}$ presents values around $\approx 3-4$. Therefore, from our data, we can regard this term as roughly constant and employ its average value, $\ln\mathcal{\bar{A}} \approx 3.7$. The resulting expression for the amplification factor is
\begin{equation}\label{ampl_approx}
    \mathcal{A} \simeq \frac{1}{2\sigma}\left[ - \ln(v_0) + \ln\left(\epsilon \sqrt{\frac{4q}{4-q}}\right) + \ln\mathcal{\bar{A}} \right] +\sqrt{\frac{4-q}{4q}}V_0\,.
\end{equation}

\subsection{Incorporating the advection of parasitic modes}

\begin{figure*}[t]
\centering
\includegraphics[width=17cm]{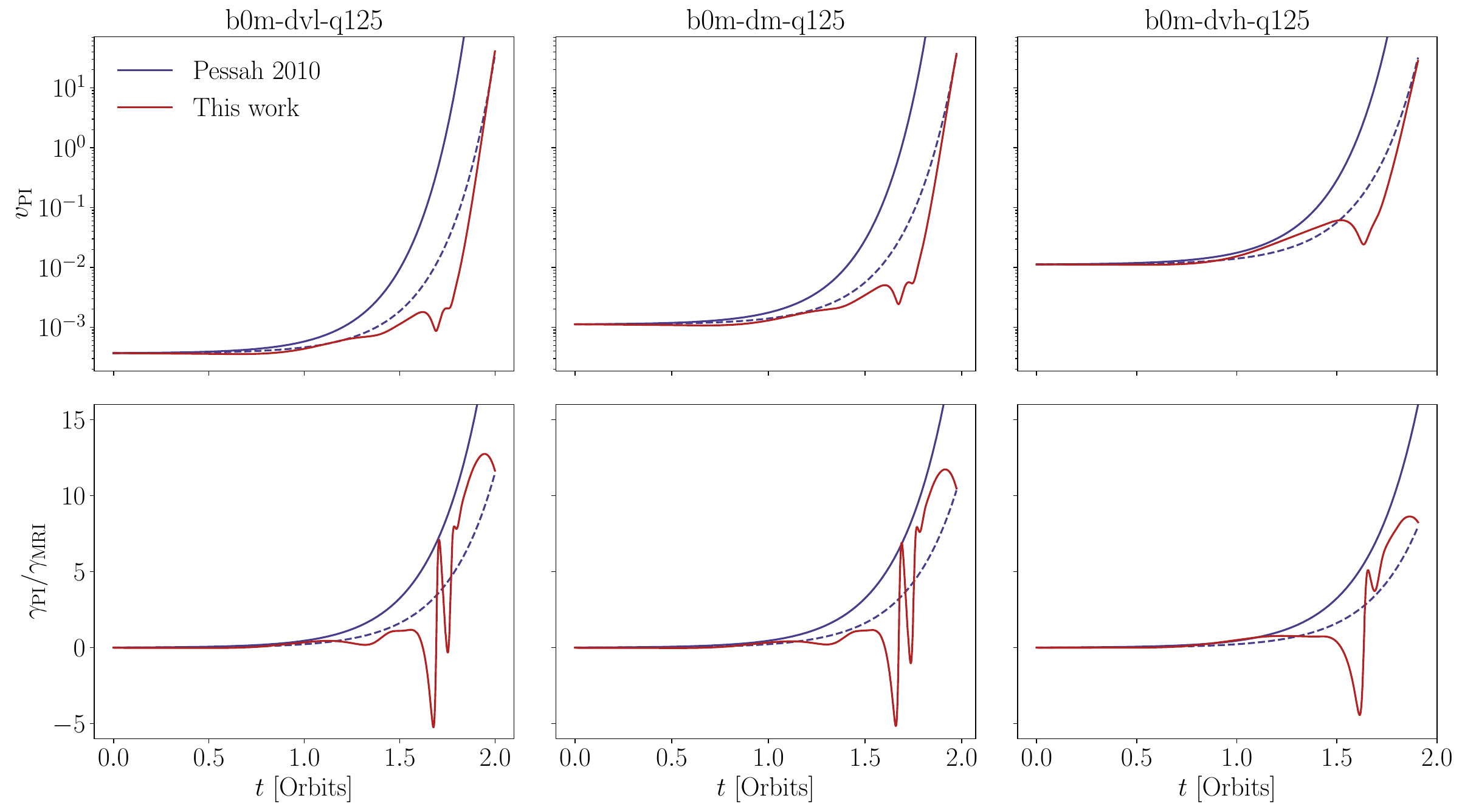}
\caption{Top panels: time evolution of the parasitic velocity for different initial amplitudes of the PIs (from left to right, $v_0 = \{0.112,1.12,11.2\}\times 10^{-3}$) and a fixed initial MRI field, $B_0 = 3.33 \times 10^{-2}$. The solid blue lines refer to the evolution of the parasitic velocity given in Eq.~\eqref{v_pi_p10}, whereas the red lines correspond to our approach. The dashed blue lines refer to the same velocity from Eq.~\eqref{v_pi_p10}, but adding the correcting factor $f$ to the parasitic growth rate from Eq.~\eqref{g_pi_pessah2010}. Bottom panels: time evolution of the normalized parasitic growth rate. Adding the correcting factor $f$ to the expression from Eq.~\eqref{g_pi_pessah2010} (dashed blue lines), the parasitic growth rate roughly agrees with the one we obtain in our study (solid red lines).}
\label{fig:effective_grate}
\end{figure*}

It can be seen in Fig.~\ref{fig:ampl_1} that the expression from~\eqref{ampl_approx} underestimates $\mathcal{A}$ by a factor $\sim 2$, compared to our results (dashed lines with respect to solid lines). This can be understood by the fact that incorporating the advection of the parasitic wavevector $\kh$ by the background shear effectively reduces the growth rate of the parasitic modes, as stated in~\cite{Latter:2010}.

Using the results from our calculations, we can compute the factor, $f$, that needs to be applied to the parasitic growth rate from Eq.~\eqref{g_pi_pessah2010} in order to match the amplification factors obtained with our approach which included the background shear. Employing all the runs from Table~\ref{tab:runs2}, we obtain that, on average, the factor needed is
\begin{equation}\label{factor_gpi}
    f = 0.498 \pm 0.006\,.
\end{equation}
We note that the value of this fraction, $f$, is found by analyzing the parasitic modes that enable the breakdown of the primary MRI mode the fastest. The result is insensitive to the initial amplitudes of either the primary MRI or the secondary parasitic modes.
The value we obtain agrees with the estimate made by~\cite{Latter:2010}, who argued a reduction by a factor $\approx 2$ by estimating the ``average'' growth rate during the time interval in which the parasites grow super-exponentially. This time interval is, in fact, short, because of the background disk shear.  In our case, we aimed to apply a correcting factor to the parasitic growth rate from Eq.~\eqref{g_pi_pessah2010} to account for the background shear. We, therefore, equate the amplitude of the non-advected parasitic velocity from Eq.~\eqref{v_pi_p10} 
near the primary MRI breakdown to the one obtained with the approach of this work (see Fig.~\ref{fig:effective_grate}). We found that the correcting factor $f$ obtained in this way also leads to roughly similar growth rates at saturation time.

The top panels in Fig.~\ref{fig:effective_grate} depict the evolution of the parasitic velocity using Eq.~\eqref{v_pi_p10} (solid blue line), using the same equation but introducing the factor $f$, so that $\gamma_{\rm PI}^{\rm P10} \rightarrow f\gamma_{\rm PI}^{\rm P10}$ in Eq.~\eqref{g_pi_pessah2010} (dashed blue line), and the actual evolution of a parasitic mode (solid red line) obtained from solving the initial value problem. As expected, the parasitic mode with a constant wavevector $\kh$  aligned with the MRI velocity (solid blue curve), as implied by Eq.~\eqref{v_pi_p10}, grows faster than the parasitic modes whose wavevector is advected by the background shear flow (solid red curve). In the bottom panels we show the evolution of the normalized parasitic growth rate. By introducing the factor $f$ from Eq.~\eqref{factor_gpi} in the analytical expression for the parasitic growth rate from~\cite{Pessah:2010}, the parasitic mode with constant wavevector $\kh$ can reach the same amplitude as the parasitic mode with time-dependent wavevector advected by the background shear.
Including this factor $f$ in Eq.~\eqref{g_pi_pessah2010} in the amplification factor we obtain
\begin{equation}\label{ampl_approx_f}
    \mathcal{A} \simeq \frac{1}{2f\sigma}\left[ - \ln(v_0) + \ln\left(\epsilon \sqrt{\frac{4q}{4-q}}\right) + \ln\mathcal{\bar{A}} \right]\,,
\end{equation}
where, supported by the results in Fig.~\ref{fig:ampl_1}, we neglect the dependence of  $\mathcal{A}$ on $V_0$. 

\subsection{An independent check using parasitic mode dynamics}

Equation~\eqref{ampl_approx_f} has been obtained in a somewhat ad hoc way by bringing together various approximations without much regard for rigor. As an independent way to check whether the dependencies implied can accurately describe the results we obtained by solving for the full dynamical evolution of the parasites feeding off exponentially growing MRI modes while being advected by the background shear, we proceed as follows. Using the results from Tables~\ref{tab:runs} and~\ref{tab:runs2}, we seek to find the coefficients in the expression:
\begin{equation}\label{ampl_coeffs}
    \mathcal{A} \simeq \frac{1}{\sigma}\left[ - b'\ln(v_0) + c'\Bigg(\ln\left(\epsilon\sqrt{\frac{4q}{4-q}}\right) + \ln\mathcal{\bar{A}}\Bigg) \right]\,.
\end{equation}
Here, we consider $\epsilon = 1$. Using the runs from Table~\ref{tab:runs} ($q = 1.5$), we perform a linear fit and find that $b' = 1.022\pm 0.014$ and $c' = 0.714 \pm 0.019$, with $R^2 = 0.996$. Alternatively, using the results from Table~\ref{tab:runs2} ($q = 1.25$), the coefficients are $b' = 1.038 \pm 0.011$ and $c' = 0.832 \pm 0.019$, and $R^2 = 0.997$. Both coefficients are, for both values of $q$, very close to unity, especially $b'$. The fact that $c'$ takes slightly smaller values than expected might be due to the term with $\ln \bar{\mathcal{A}}$.  In any case, the fact that all coefficients are order $\mathcal{O}(1)$ shows that the dependencies in Eq.~\eqref{ampl_approx_f} are indeed roughly correct.

\subsection{A simple expression for the amplification factor}

With all these considerations we can provide a simple expression that encapsulates the key processes involved and make use of the insights provided by our calculations to provide a reasonable accurate description of the amplification of the MRI as limited by the fastest parasitic modes as
\begin{equation}\label{ampl_approx_rough}
    B_0(t_{\rm sat}) \simeq \frac{\bar{B}_z}{0.135} \left[ - \ln(v_0) + \ln\left(\sqrt{\frac{4q}{4-q}}\right) + 3.7 \right]\,,
\end{equation}
where $\bar{B}_z$ is the large-scale vertical magnetic field, $v_0$ is the initial amplitude of the parasitic mode, and $q$ is the shear parameter.

\section{Comparison with numerical simulations}
\label{sec::comparison}

\subsection{Amplification factors}

The role of PIs in the breakdown of the primary MRI mode has been studied through numerical simulations thoroughly during the last decade~\citep{Obergaulinger:2009,Lesur:2011,Sorathia:2012,Murphy:2015,Rembiasz:2016a,Hirai:2018,Gogi:2018}. The use of numerical simulations has allowed the nonlinear regime of the MRI to be studied. However, numerical studies have not usually focused on addressing the role that seed perturbations, that excite unstable modes (either the MRI itself or the parasites) may play on the saturation level of the ensuing turbulence.

\cite{Rembiasz:2016b} addressed this issue by carrying out shearing box simulations using the pseudo-spectral code \textsc{SNOOPY}~\citep{Lesur:2005,Lesur:2007} for different values of the initial amplitudes to seed the MRI, $V_0$, and the PIs, $v_0$.  Figure~\ref{fig:ampl_1} shows the amplification factors they obtained with black triangles. In the left panel, the initial parasitic amplitude is fixed at $v_0 = 1.12 \times 10^{-2}$, whereas in the right panel the initial MRI velocity is fixed at $V_0 = 2.2\times 10^{-3}$. Their results show a steeper dependence on the initial MRI velocity (left panel) and on the initial parasitic amplitudes (right panel). These trends are captured by the authors proposing that the amplification factor scales according to
\begin{equation}\label{fit_sims}
    \mathcal{A}(V_0,v_0) = a\ln V_0 + b\ln v_0+c\,,
\end{equation}
where $a = 5.4 \pm 0.55$, $b=-20.2\pm 1.2$, $c = -101\pm 13$. The value of the factor $c$ differs from the one from the fit in Fig.~\ref{fig:ampl_1} because the units employed in~\cite{Rembiasz:2016b} are different.

We attribute the discrepancy in the amplification factors we obtained with respect to those obtained in numerical simulations by \cite{Rembiasz:2016b}, shown in Fig.~\ref{fig:ampl_1}, to the different ways the modes involved are excited. We speculate, on rather reasonable grounds, that the larger amplification factors seen in numerical simulations could be due to the fact that the initial conditions do not seed directly the fastest modes involved.
\cite{Rembiasz:2016b} only excited the fastest growing velocity MRI field (letting the magnetic MRI field grow later), and they excited a large set of parasitic velocities applying random factors to their initial amplitudes. 
In our approach, only the fastest primary (MRI) and the secondary (parasitic) modes interact to reach saturation. The initial primary modes correspond to the fastest MRI eigenmode in Eq.~\eqref{mri_fields}. In addition, every time we seed secondary perturbations, we do so by exciting parasitic eigenmodes at the same initial time we allow the MRI mode to start evolving. Under these conditions, we define the amplitude of the MRI at saturation by equating the (kinetic) energy densities of the fastest MRI mode and the fastest parasitic mode. 
For initial conditions that exclusively involve perturbations in the velocity amplitude of the primary, $V_0$, it takes longer for the primary MRI mode, with the corresponding $B_0$, to emerge and grow exponentially at the fastest rate. This produces a delay in the onset of the secondary mode that grows to saturation, resulting in a higher amplitude for the MRI mode near breakdown that is higher than the one we would have obtained had the primary MRI mode been seeded directly. This may explain the trend exhibited by numerical results on the left panel of Fig.~\ref{fig:ampl_1}. 
Concerning the discrepancies observed in terms of the initial amplitude of the secondary perturbations, on the right panel of Fig.~\ref{fig:ampl_1}, it seems natural for saturation to be reached at a lower amplification when the fastest secondary mode is excited directly. This is because depositing the same amount of kinetic energy in a broad spectrum of secondary velocity perturbations is less efficient in seeding the parasites, allowing the primary to reach a higher amplitude.

It is worth mentioning that \cite{Rembiasz:2016b} performed simulations with another numerical code, \textsc{Aenus}~\citep{Obergaulinger:2008}, with different radial boundary conditions, physical assumptions and numerical schemes. The resulting amplification factor differed by a factor 5 from that obtained with the \textsc{SNOOPY} code ($\mathcal{A} \approx 19$ with \textsc{Aenus} instead of $\mathcal{A} \approx 90$ with \textsc{SNOOPY}). Moreover, they found that employing a different form of the initial perturbations also changed the amplification factor, obtaining $\mathcal{A} \approx 60$ instead of $\mathcal{A} \approx 90$. Thus, differences in the simulation setup and the use of different numerical codes can have an impact on the amplification factor. 

\subsection{Other aspects}

Numerical studies have also shown that the PIs eventually grow super-exponentially, reaching an amplitude comparable to the MRI and leading to a turbulent regime. It has also been found that, in the ideal MHD case, the fastest growing parasitic mode is aligned with the MRI velocity field~\citep{Rembiasz:2016a,Hirai:2018}, which is in agreement with our findings. However, the parasitic wavenumbers at saturation found in numerical simulations~\citep{Rembiasz:2016a,Hirai:2018} are somewhat larger than those obtained here. This can be explained by the existence of other MRI modes (that grow more slowly), which results in a shear flow that is not purely sinusoidal. Moreover, the interaction of the PIs with the channels can cause the layered structure to become narrower, which induces a smaller parasitic mode just before the primary MRI mode breakdown~\citep{Hirai:2018}. These nonlinear effects are not captured by our approach.

Incorporating the advection of the parasitic modes has allowed us to understand that the modes that end up being able to reach amplitudes comparable to the MRI quite generically have initial horizontal wavelengths that are a factor of a few smaller than those previously inferred. We found that the fastest growth occurs when the parasitic wavevector is aligned with the MRI velocity field, in agreement with previous analytical and numerical studies. However, these modes must be properly resolved throughout their evolution in order for numerical simulations to capture their dynamics, and thus the MRI saturation, faithfully.

Using more than ten grid cells per wavelength is usually assumed to be enough to properly resolve the MRI~\citep{Rembiasz:2016a}. It is clear from our findings that it is equally important to resolve the evolution of the parasitic wavevector, as it is advected by the background shear flow. By performing box simulations with different aspect ratios and grid resolution,~\cite{Rembiasz:2016a} found that at least 60 zones per MRI channel are needed to obtain convergent results. This is due to the fact that the KH-type parasitic modes, triggered by the shear layer between MRI channels, develop finer spatial structures that need to be resolved. As a result, an initially leading wave with large (negative) $k_x(0)$ may be numerically damped early on, and thus reach positive values of $k_x(t)$ at a significantly lower amplitude than initially, which would complicate any estimate made from it. This issue can be explained by our results. The fastest parasitic modes have been found to be initially $\sim 3-5$ times smaller than the MRI channels, meaning that an increased resolution in the horizontal plane might be needed to properly capture both the MRI and the PIs from the beginning of the simulation. If resolved, all the simulations with different numerical resolution should lead to the same amplification factor. 

\section{Summary, discussion, and implications}
\label{sec::conclusions}

We have investigated the saturation of the MRI via parasitic modes relaxing important approximations previously invoked. This allowed us to obtain the most accurate calculation of the amplification factor to which the MRI can grow until the parasitic modes reach a comparable amplitude. To accomplish this, we carried out the first systematic analysis of the evolution of the PIs considering the temporal dependence that arises from the advection of the secondary parasitic modes feeding off exponentially growing primary MRI modes. 
Our approach involved solving a large number of initial value problems providing initial amplitudes for the modes involved. Even though the values for the initial amplitudes spanned several orders of magnitude we found that the amplification factor incurred by the MRI is remarkably robust, depending only logarithmically on the initial amplitude of the parasitic modes. This is overall in reasonable  agreement, within a factor of two, with the amplification factors found in numerical simulations~\citep[e.g.,][]{Hawley:1995,Obergaulinger:2009,Murphy:2015,Rembiasz:2016a,Rembiasz:2016b}.

Building on previous work that has led to physical intuition on the processes involved in the breakdown of the primary MRI mode via PIs, and guided by our numerical results, we have provided a simple analytical expression in Eq.~\eqref{ampl_approx_rough} that describes quite reasonably well the amplification of magnetic fields driven by the MRI. 

The discrepancies with numerical studies mentioned above may highlight interesting issues with simulations. All in all, the amplification factors implied by our results differ by about a factor $\approx 2$  from those reported in~\cite{Rembiasz:2016b}. The slope on the right panel of Fig.~\ref{fig:ampl_1} is also considerably larger than the one we obtain, although the trends are similar. The discrepancy between our results and those in~\cite{Rembiasz:2016b} can be partly explained by the different ways in which both the MRI and the parasites are excited in each of these settings (see Sect.~\ref{sec::comparison} for a more detailed discussion). Another potential reason for which our predictions for saturation differ from numerical results may be because our approach does not capture the nonlinear interaction between primary and secondary that may arise near termination ~\citep[see, e.g.,][]{Murphy:2015,Gogi:2018}. In the nonlinear regime of the KH instability, a growth rate reduction is expected when the primary MRI mode breakdown approaches, so our predictions might underestimate the amplification factor~\citep[cf.][]{Keppens:1999,Obergaulinger:2010}. The plausible reasons we offer to explain the difference in amplification factors obtained via numerical simulations and via our approach can be explicitly tested by seeding numerical simulations exciting the fastest primary and secondaries.

In summary, with all the caveats stated, we have built an effective model to compute the saturation amplitude of the MRI, which differs up to a factor of roughly 2 from numerical results, which is smaller than the spread in the values provided by different codes using different initial conditions. This is the first approach that can achieve this and also highlights which parameters play the most relevant role in determining MRI saturation.

As shown in Eq.~\eqref{ampl_fact_general}, the amplification factor is expressed in terms of the volume-averaged Maxwell stress tensor, $\bar{\mathcal{M}}_{xy}$. 
Thus, the effective model we have built here can be employed to predict the value at saturation of the turbulent stress tensors, since the Reynolds stress, $\bar{\mathcal{R}}_{xy}$, can be found via~\citep{Pessah:2006a}
\begin{equation}\label{reynolds}
    \bar{\mathcal{R}}_{xy} = -\frac{q}{4-q}\bar{\mathcal{M}}_{xy}\,.
\end{equation}
These components of the Maxwell and Reynolds stresses determine the radial flux of angular momentum via the total stress
\begin{equation}\label{ang_mom_flux}
    \bar{\mathcal{T}}_{xy}  \equiv \bar{\mathcal{R}}_{xy}-\bar{\mathcal{M}}_{xy} \,.
\end{equation}

Numerical simulations of astrophysical systems need to account for many physical processes (e.g., realistic equations of state, neutrino and radiative processes, etc.), usually limiting their spatial resolution. For this reason, and in order to resolve the initial field amplification due to the MRI, several works assume initial large-scale magnetic fields with unrealistically large magnitudes~\citep[e.g.,][]{Etienne:2012,Gold:2014,Ruiz:2018,Ruiz:2021,Bamber:2024}. Our effective model may be regarded as a means to predict the final magnetic field amplitude when secondary instabilities limit the MRI growth. This presents an alternative to invoking the large initial magnetic fields usually assumed in order to resolve the MRI.

Our findings may help develop more realistic effective models for MRI-driven turbulence to go beyond current approaches that employ effective viscosity terms~\citep{Fernandez:2013,Shibata:2017,Fujibayashi:2018,Fujibayashi:2020,Radice:2020,Just:2023}, which are based on the alpha-viscosity prescription from~\cite{Shakura:1973}.
The resulting radial flux of the angular momentum [cf.~Eq.~\eqref{ang_mom_flux}] can be used as a viscosity term in the momentum equation that allows transporting angular momentum at the correct rate even though the MRI is not well-resolved by the numerical simulation. 
The functional dependence of the effective model from Eq.~\eqref{ampl_approx_rough} on the initial amplitudes of the modes and on the shear parameter $q$ improves its adaptability to different disk conditions with respect to the widely used alpha-viscosity\footnote{We refer the reader to~\cite{Pessah:2008a} for further details.}. 
For example, the subgrid model presented in~\cite{Miravet:2022} makes use of the findings made by~\cite{Pessah:2010} to build  evolution equations for the turbulent kinetic energy densities of both the MRI and PIs. 
This model can be improved by incorporating the findings of this work,  namely, by including the factor $f \simeq 1/2$ from Eq.~\eqref{factor_gpi} in the parasitic growth rate that appears in these equations. 

The effective model for MRI saturation we obtained can also be used to develop closure relations to link mean magnetic fields with turbulent stresses which are critical in models that go beyond viscous prescriptions for angular momentum transport~\citep[e.g.,][]{Kato:1993, Kato:1995, Ogilvie:2003, Pessah:2006b, Pessah:2008a}. This is a sensible approach, 
because even though nonlinear effects may result in some amplitude fluctuations during the turbulent state, the effective model presented here can provide an estimate of the average MRI amplitude after saturation ~\citep[e.g.,][]{Rembiasz:2016b,Hirai:2018,Reboul-Salze:2021}.

The insights obtained in better understanding the saturation mechanism for the MRI have the potential to bring us a step closer to developing more realistic dynamical mean-field dynamo models where MRI-driven turbulence is naturally built-in~\citep{Gressel:2010, Gressel:2015, Gressel:2022, Vourellis:2021}.

\begin{acknowledgements}
    
We thank Pablo Cerdá-Durán, José A.~Font, Henrik Latter, George Mamatsashvili, Kaushik Satapathy, and Loren E.~Held for helpful discussions and useful comments. 
We also thank the referee for a comprehensive and useful report that helped us improve the manuscript. 
MMT acknowledges support from the Ministerio de Ciencia, Innovación y Universidades del Gobierno de España through the
``Ayuda para la Formación de Profesorado Universitario'' (FPU) fellowship No.~FPU19/01750 and the ``Ayuda FPU Complementaria de Movilidad para Estancias Breves en Centros Extranjeros'' fellowship No.~EST23/00420, from the Spanish Agencia Estatal de Investigación (grant PID2021-125485NB-C21) funded by MCIN/AEI/10.13039/501100011033 and ERDF A way of making Europe, and from the Science and Technology Facilities Council (STFC), via grant No.~ST/Y000811/1. MEP gratefully acknowledges support from the Independent Research Fund Denmark via grant ID 10.46540/3103-00205B.

This work has used the following open-source packages: \textsc{NumPy}~\citep{harris:2020}, \textsc{SciPy}~\citep{scipy:2020} and \textsc{Matplotlib}~\citep{Hunter:2007}.
\end{acknowledgements}

\bibliographystyle{aa}
\bibliography{draft}

\begin{appendix}

\section{Evolving the system of equations for the parasitic instabilities}\label{app:time_ev}

Having obtained the initial values of the Fourier amplitudes of the velocity and magnetic fields of the PIs, we can evolve the system of Eqs.~\eqref{alpha_eq} and~\eqref{beta_eq}. 
To accomplish this we write it in vector form as
\begin{equation}\label{coupled_eqs}
    \partial_t \boldsymbol{x}(t) = \boldsymbol{A}(t)\boldsymbol{x}(t)\,,
\end{equation}
where the vectors $\boldsymbol{x}$ contain
the Fourier amplitudes $\alpha_n(t)$ and $\beta_n(t)$ and 
$\boldsymbol{A}(t)$ is the matrix of coefficients.
This coupled linear differential equation system can be decoupled, and solved, as follows. 

Let the matrix $\boldsymbol{A}(t)$ be expressed in terms of a diagonal matrix, $\boldsymbol{\Lambda}(t)$:
\begin{equation}\label{diag_matrix}
    \boldsymbol{\Lambda} = \boldsymbol{S}^{-1}\boldsymbol{A}\boldsymbol{S}\,,
\end{equation}
where $\boldsymbol{S}$ is a matrix built with the eigenvectors of $\boldsymbol{A}(t)$ as columns and $\boldsymbol{\Lambda}$ is a diagonal matrix with the eigenvalues of $\boldsymbol{A}$ as elements. 
The equations of motion~\eqref{coupled_eqs} can be written in terms of the variable
\begin{equation}
    \boldsymbol{\eta} = \boldsymbol{S}^{-1}\boldsymbol{x}\,,
\end{equation}
as a set of decoupled equations
\begin{equation}\label{decouple}
    \partial_t \boldsymbol{\eta} (t) = \boldsymbol{\Lambda} \boldsymbol{\eta}(t)\,,
\end{equation}
which can be cast independently for each component
\begin{equation}
    \partial_t \eta_i = \lambda_i \eta_i\,.
\end{equation}
This is valid only when $\boldsymbol{S} \neq \boldsymbol{S}(t)$, and we assume this is the case in a small enough time interval $\Delta t$. The solutions can now be trivially obtained as
\begin{equation}
    \eta_i(t) = \exp\left[{\int_{t-\Delta t}^{t}\lambda_i(\tau) d\tau}\right] \, \eta_i(t-\Delta t)\,.
\end{equation}
We can define the matrix composed of elements given by the right-hand-side of the above equation:
\begin{equation}
    E_{ij} \equiv \delta_{ij} \exp \left[
    \int_{t-\Delta t}^{t} \lambda_i(\tau) \, d\tau 
   \right]\,.
\end{equation}
For small enough time intervals $\Delta t$, we can assume that $\lambda_i$ are constant\footnote{Of course $\lambda_i$ can change value between consecutive time intervals.} to obtain 
\begin{equation}\label{approx_lambda}
    E_{ij} \approx \delta_{ij} e^{\Delta t\lambda_i }\,.
\end{equation}
In terms of the matrix $\boldsymbol{E}$, the solutions we seek satisfy
\begin{equation}
    \boldsymbol{x}(t) = \boldsymbol{S}\boldsymbol{E}\boldsymbol{S}^{-1}\boldsymbol{x}(t-\Delta t)\,.
\end{equation}

In solving the equations given by~\eqref{decouple}, we employ the approach shown in Eq.~\eqref{approx_lambda} using an adaptive timestep $\Delta t$ that decreases as $\gamma_{\rm PI}$ increases, as shown in Table~\ref{tab:delta_t}. The initial timestep is set to $\Delta t_0 = 0.1$.

\begin{table}[h]
\caption{Timestep employed in terms of the parasitic mode growth rate.}
    \centering
    \begin{tabular}{|c|c|}
        \hline
       $\gamma_{\rm PI}/\gamma_{\rm MRI}$ & $\Delta t$/$\Delta t_0$ \\  
       \hline
       $< 0.1$ & 1 \\
       $\geq 0.1$ & $10^{-1}$  \\
       $\geq 1$ &  $10^{-2}$  \\
       $\geq 25$ & $10^{-3}$  \\
     \hline
\end{tabular}
\vspace{2mm}
    \label{tab:delta_t}
\end{table}

\section{Physical structure of the parasitic modes}\label{app:phys_structure}

To study the structure of the parasitic modes and the disruption of the MRI channels, we calculate the components of the vorticity and current density perpendicular to the plane ($\boldsymbol{\check{k}_{\rm h}},\check{\bf{z}}$):
\begin{align}\label{vort_curr}
    \delta \omega_{\perp} (t;\boldsymbol{x}) & = (\curl \boldsymbol{v})\cdot \boldsymbol{\check{k}_{\rm p}} \,, \\
    \delta j_{\perp} (t;\boldsymbol{x}) & = (\curl \boldsymbol{b})\cdot \boldsymbol{\check{k}_{\rm p}}\,,
\end{align}
where $\boldsymbol{\check{k}_{\rm p}}$ is the direction perpendicular to ($\boldsymbol{\check{k}_{\rm h}},\check{\bf{z}}$): $\boldsymbol{\check{k}_{\rm p}} \equiv \check{\bf{z}}\wedge \boldsymbol{\check{k}_{\rm h}}$. Using Eqs.~\eqref{v_fou} and~\eqref{b_fou}, we obtain
\begin{align}\label{vort_curr2}
    \delta \omega_{\perp} (t;\boldsymbol{x}) & = -\frac{i}{k_{\rm h}} \suma [k_{\rm h}^2+(n+k_z)^2]\alpha_n(t)\efouv\expoh \,, \\
    \delta j_{\perp} (t;\boldsymbol{x}) & = -\frac{i}{k_{\rm h}} \suma [k_{\rm h}^2+(n+k_z)^2]\beta_n(t)\efouv\expoh \,,
\end{align}
where $k_{\rm h}h = \boldsymbol{k_{\rm h}}\cdot\boldsymbol{x}$. The total vorticity and current can be obtained by adding the contribution from the MRI fields projected onto the direction $\boldsymbol{\check{k}_{\rm h}}$: 
\begin{align}\label{vort_curr_tot}
    \omega_{\perp} (t;\boldsymbol{x}) & = V_0(t)\cos(Kz)\cos(\theta-\theta_{\rm V})+\delta \omega_{\perp} (t;\boldsymbol{x}) \,, \\
    j_{\perp} (t;\boldsymbol{x}) & = -B_0(t)\sin(Kz)\cos(\theta-\theta_{\rm B})+\delta j_{\perp} (t;\boldsymbol{x}) \,.
\end{align}

\begin{figure*}[h]
\centering
\includegraphics[width=0.23\textwidth]{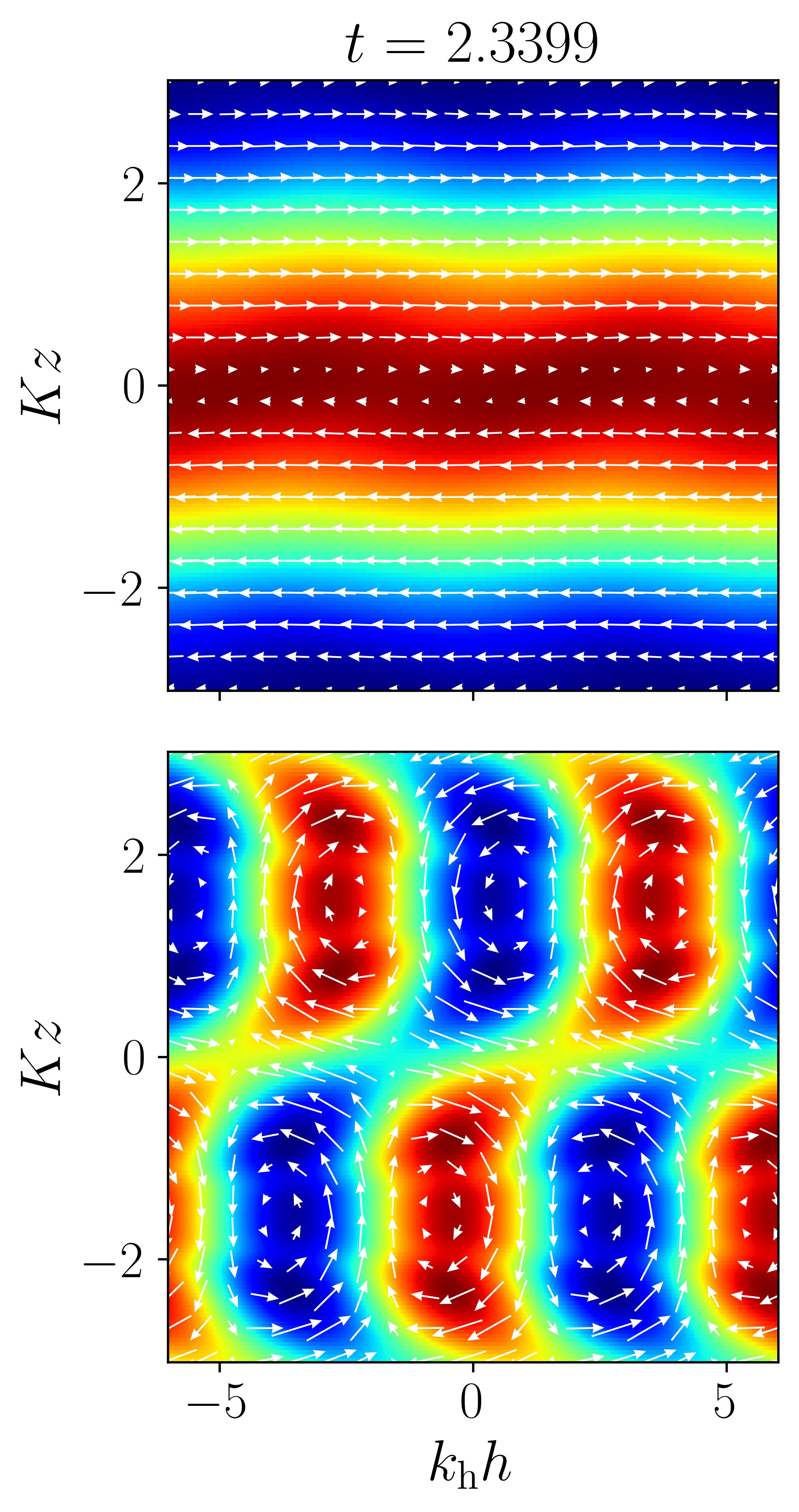}
\includegraphics[width=0.23\textwidth]{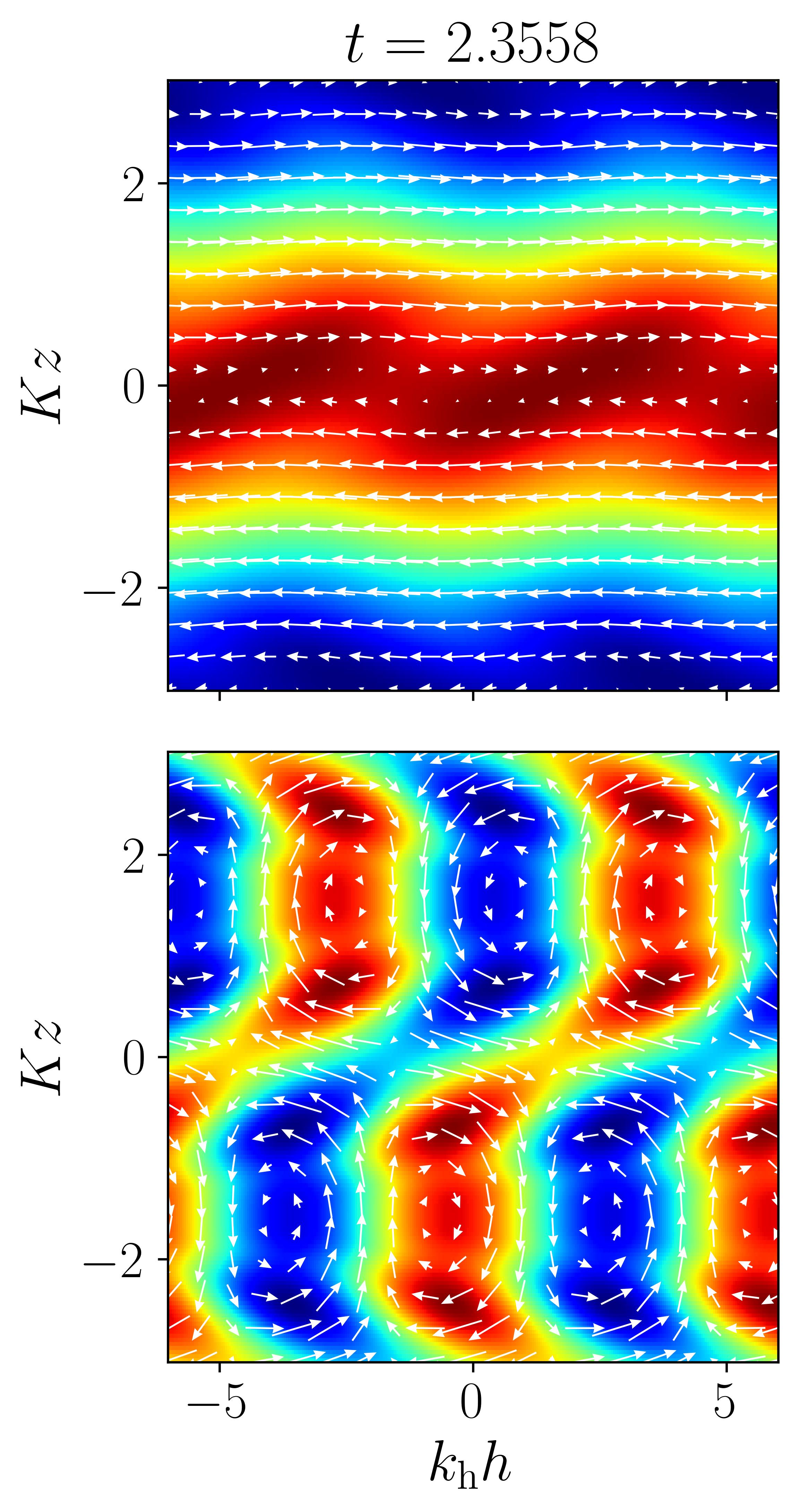}
\includegraphics[width=0.23\textwidth]{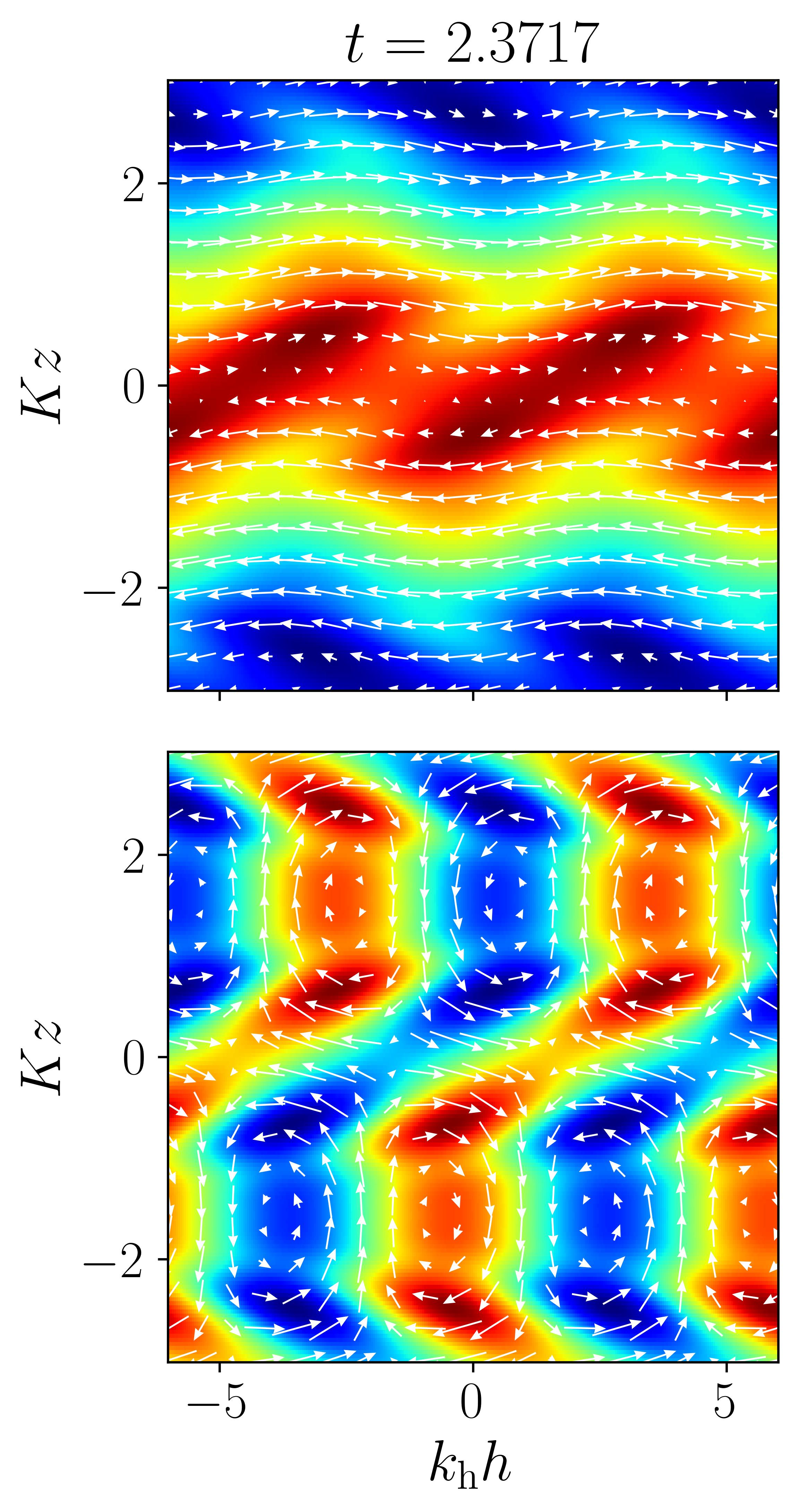}
\includegraphics[width=0.23\textwidth]{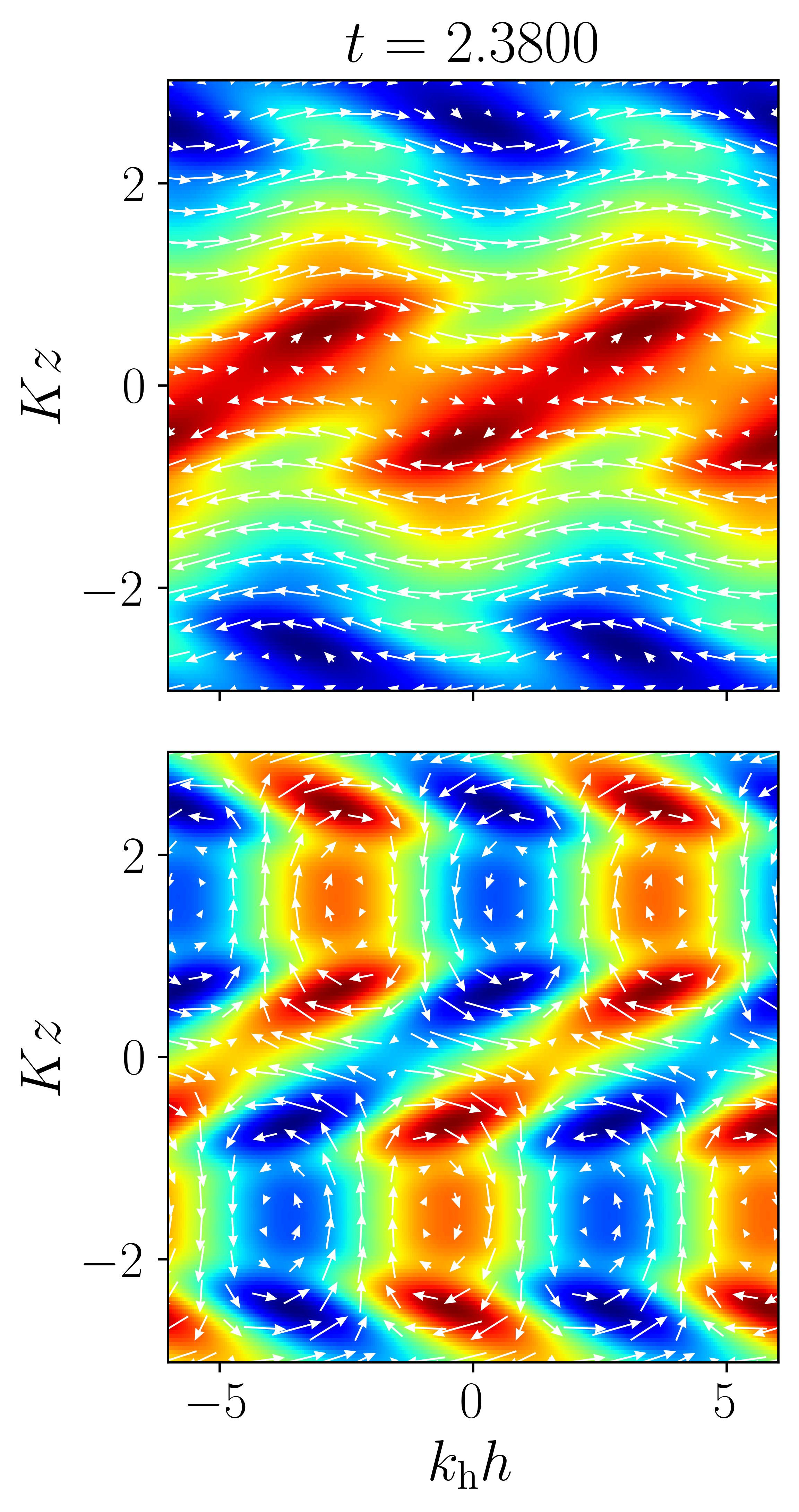}
\caption{Physical structure of the fastest parasitic modes, including the velocity field of the primary MRI mode, for different times (in terms of the number of orbits, increasing from left to right). The arrows in the top and bottom panels correspond, respectively, to the projections of the total velocity, $\boldsymbol{V}(z)+\boldsymbol{v}(h,z)$, and the parasitic velocity, $\boldsymbol{v}(h,z)$, onto the time-dependent plane ($\boldsymbol{\check{k}_{\rm h}},\check{\bf{z}}$). The color contours correspond to the associated total vorticity $\omega_{\perp}$ and the parasitic vorticity $\delta \omega_{\perp}$ projected onto the direction perpendicular to $\boldsymbol{\check{k}_{\rm h}}$.}
\label{fig:v_contours}
\end{figure*}

\begin{figure*}[h]
\centering
\includegraphics[width=0.23\textwidth]{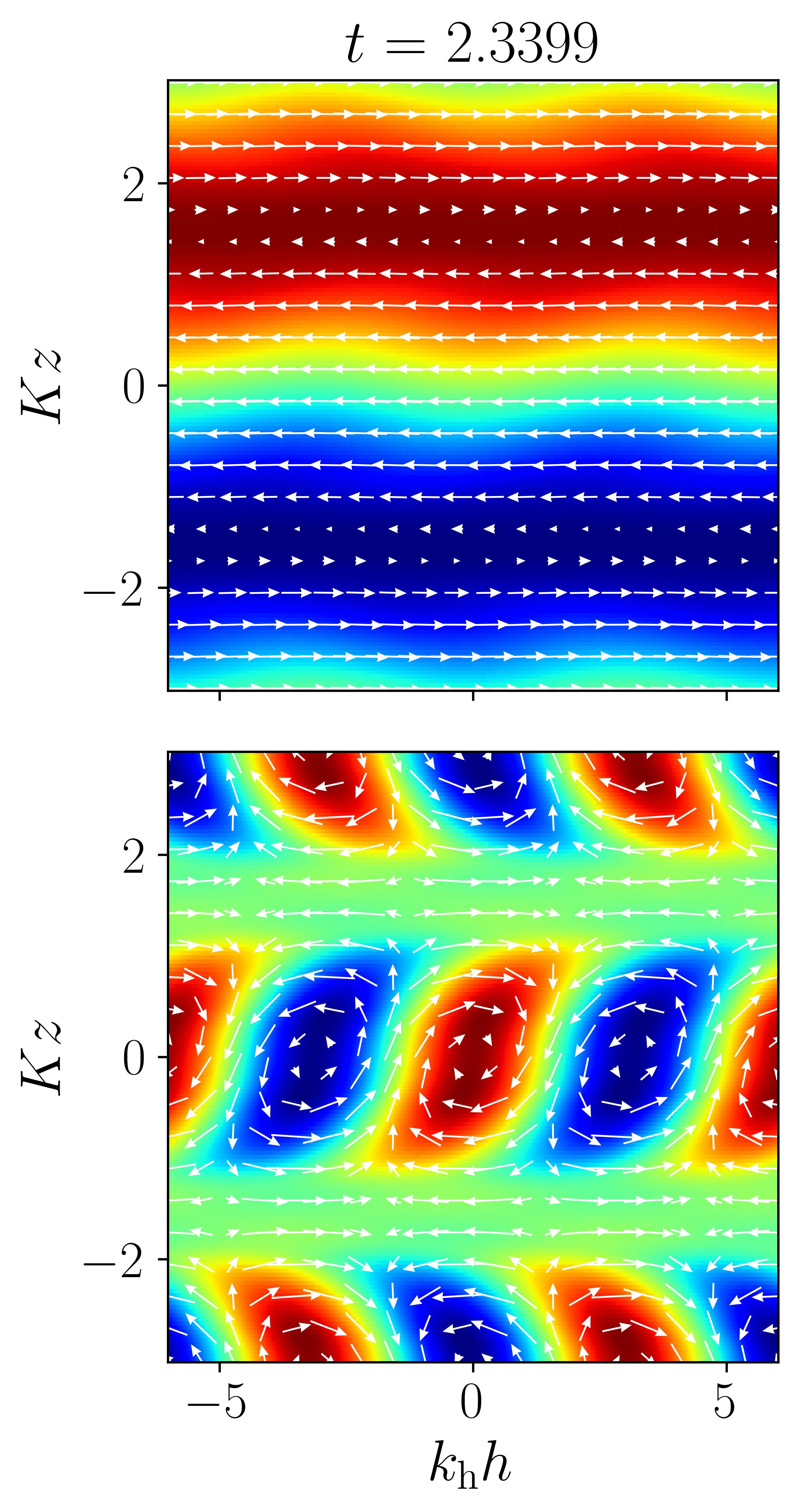}
\includegraphics[width=0.23\textwidth]{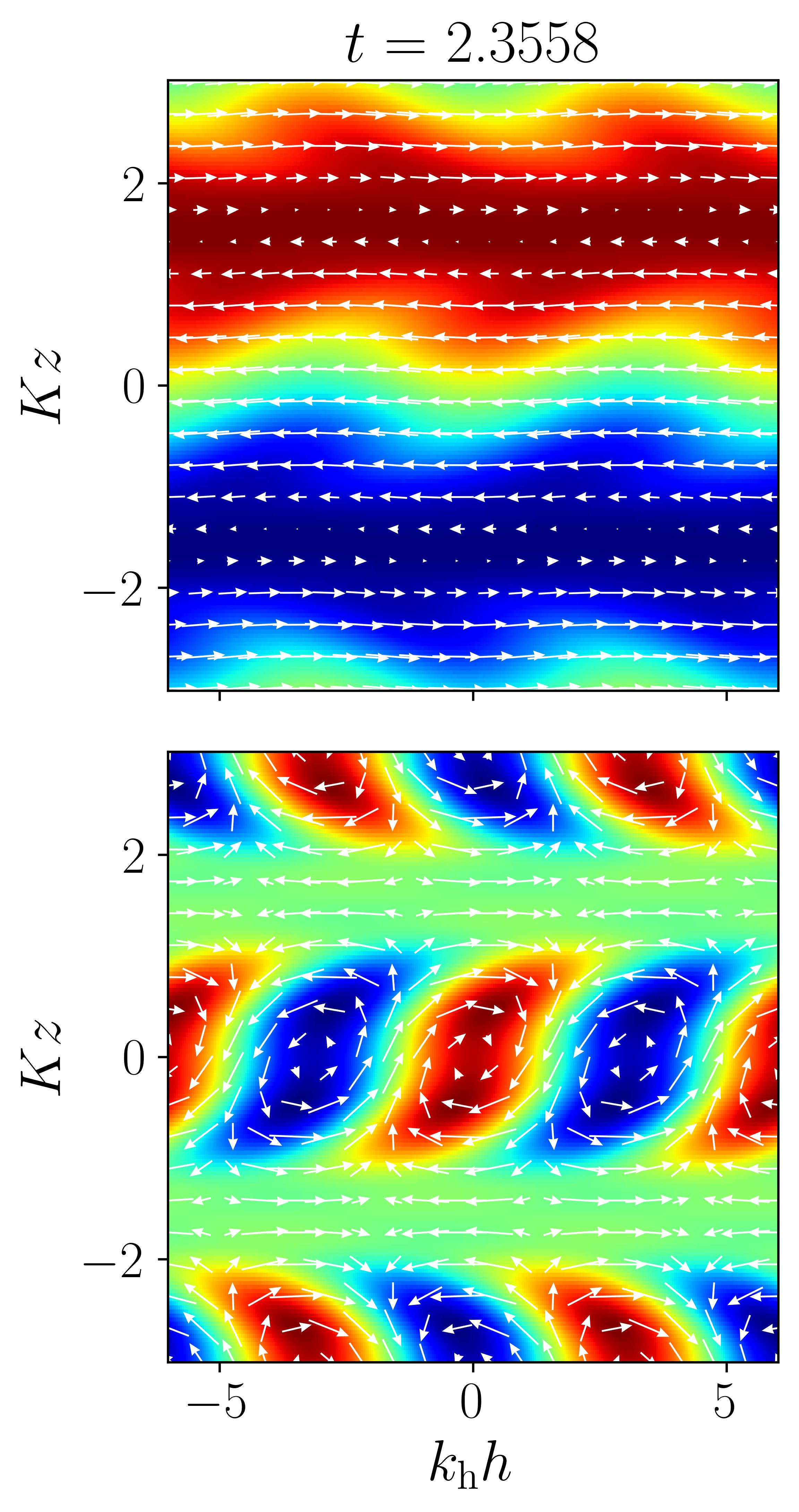}
\includegraphics[width=0.23\textwidth]{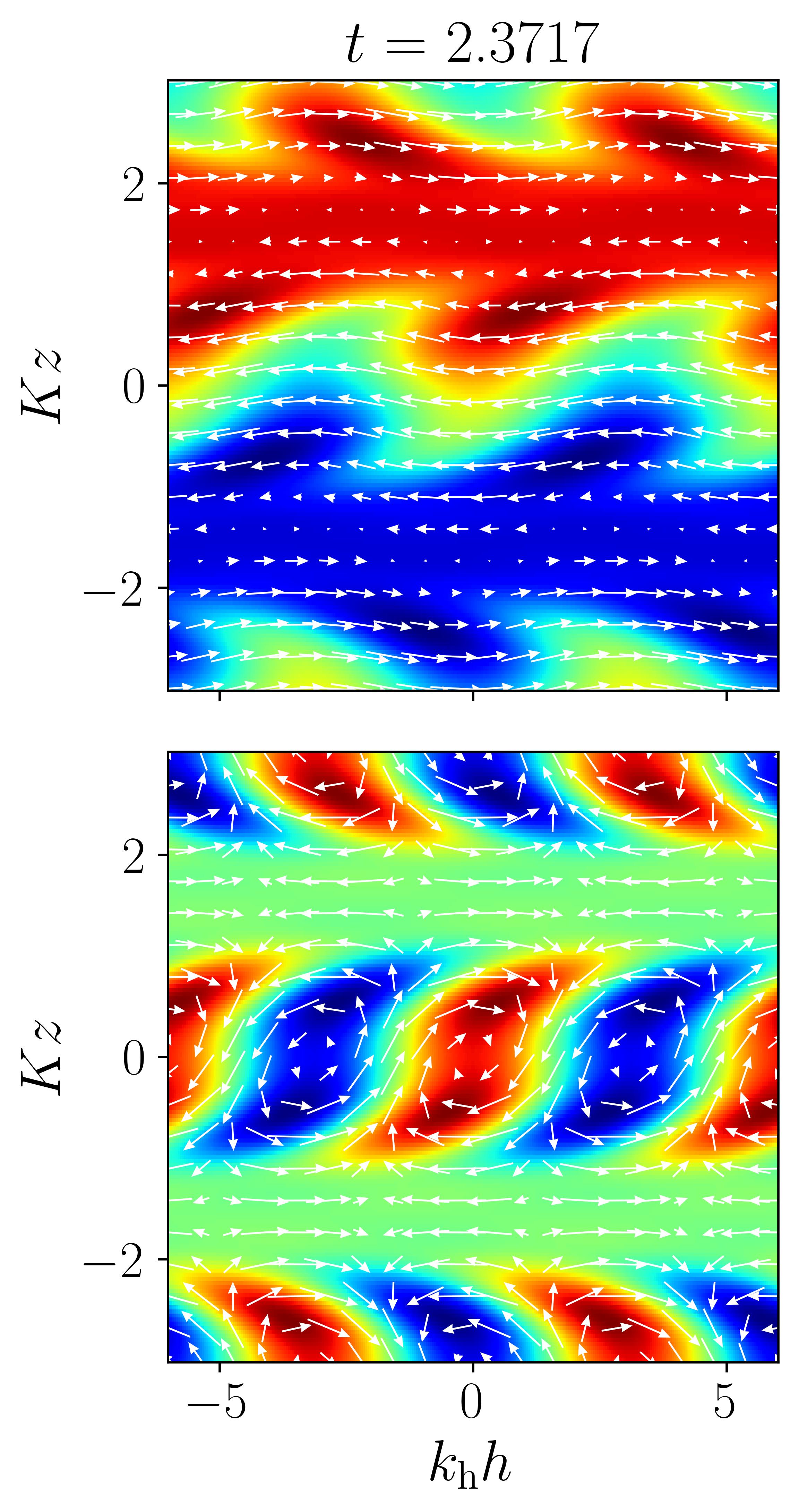}
\includegraphics[width=0.23\textwidth]{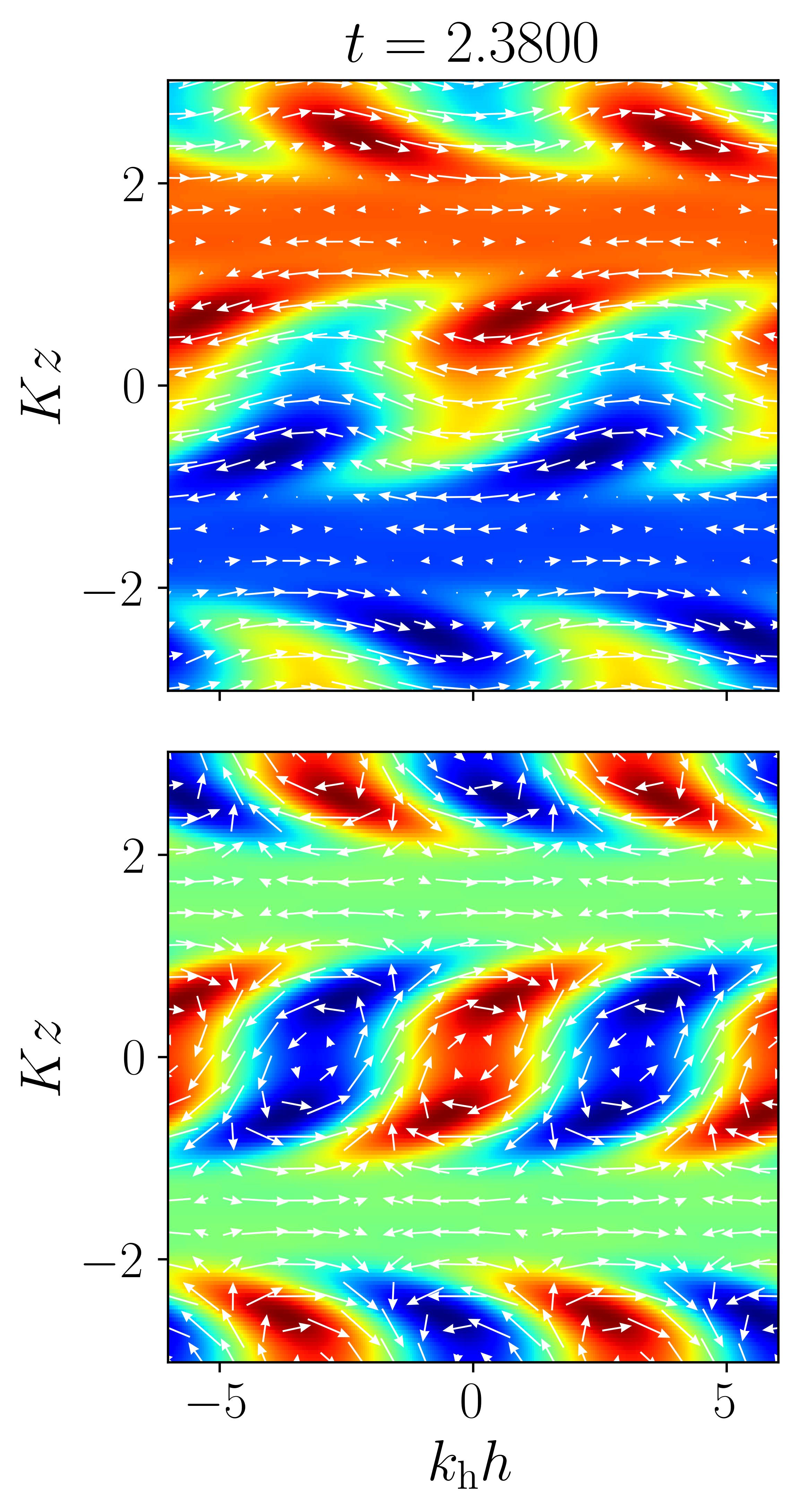}
\caption{Same as Fig.~\ref{fig:v_contours}, but including the magnetic and the current density fields instead of the velocity and the vorticity, respectively.}
\label{fig:b_contours}
\end{figure*}

In Fig.~\ref{fig:v_contours} we depict the total vorticity field $ \omega_{\perp}$ projected onto the plane ($\boldsymbol{\check{k}_{\rm h}},\check{\bf{z}}$), where the arrows represent the total velocity $\boldsymbol{V}+\boldsymbol{v}$ (top panels) and the vorticity of the parasitic mode $\delta \omega_{\perp}$ with the velocity field $\boldsymbol{v}$ (bottom panels), for different times. At early times (first panel), the MRI channels remain steady, since the fastest parasitic mode has not reached enough amplitude yet. When the primary MRI mode breakdown approaches (last three panels), the channels are disrupted by the parasitic mode, showing the familiar wave-like structure expected from KH instability modes in a periodic background. The bottom panels representing the vorticity and velocity fields of the parasitic mode show a periodic structure of equidistant vortex sheets, similar to those in Fig.~8 in~\cite{Pessah:2010}, but gradually evolving to more elongated structures. 

Alternatively, Fig.~\ref{fig:b_contours} shows the current density and magnetic fields instead of the vorticity and the velocity fields. The magnetic parasitic field again forms periodic vortex sheets that become horizontally elongated as saturation approaches. The MRI channels get disrupted and adopt the same structure as shown in Fig.~10 from~\cite{Pessah:2010}.  

\FloatBarrier

\section{Summary of the evolution runs for the parasitic modes}\label{app:runs}

We provide tables listing the parameters and important properties of the fastest parasitic modes evolved in this work:
 \begin{itemize}
     \item Table~\ref{tab:runs} summarizes several properties of the  fastest growing parasitic modes at $t=0$ and at saturation time, $t_{\rm sat}$, for different initial amplitudes of primary MRI and secondary modes. The shear factor used for these runs is $q = 1.5$.

     \item Table~\ref{tab:runs2} shows the same quantities as Table~\ref{tab:runs}, but the value of the shear parameter corresponds to $q = 1.25$, and different initial amplitudes are also used. 
 \end{itemize}

\begin{table*}[h]
\caption{Summary of the fastest parasitic modes for different initial amplitudes of the MRI modes and parasitic velocities.}
    \centering
    \begin{tabular}{|c|c|c|c|c|c|c|c|c|c|}
        \hline
       NAME & $ B_0[10^{-4}]$ & $v_0[10^{-4}]$ & $q$ & $k_{\rm h}^{\rm ini}$ & $\theta^{\rm ini}$ $[^{\circ}]$ & $k_{\rm h}^{\rm sat}$ & $\theta^{\rm sat}$ $[^{\circ}]$ & $t_{\rm sat} $ [Orbits] & $\mathcal{A}$  \\  
       \hline
        \texttt{b0vl-dvl} & $1$ & $0.1$ & 1.5 & 4.25 & 177.64 & 0.65 & 15.70 & 2.954 & 55.63  \\
        \texttt{b0vl-dl} & $1$ & $0.5$ & 1.5 & 4.50 & 177.61 & 0.70 & 15.57 & 2.927 & 48.93  \\
        \texttt{b0vl-dm}  & $1$ & $1$ & 1.5 & 4.50 & 177.61 & 0.68 & 16.07 & 2.915 & 46.15  \\
        \texttt{b0vl-dh} & $1$ & $5$ & 1.5 & 4.75 & 177.59 & 0.72 & 16.25 & 2.884 & 39.93  \\
        \texttt{b0vl-dvh} & $1$ & $10$ & 1.5 & 4.75 & 177.59 & 0.69 & 16.90 & 2.869 & 37.24 \\
        \texttt{b0l-dvl} & $5$ & $0.1$ & 1.5 & 4.88 & 177.36 & 0.70 & 18.69 & 2.613 & 55.54  \\
        \texttt{b0l-dl} & $5$ & $0.5$ & 1.5 & 3.88 & 177.23 & 0.72 & 15.06 & 2.587 & 49.26  \\
        \texttt{b0l-dm} & $5$ & $1$ & 1.5 & 3.88 & 177.23 & 0.70 & 15.57 & 2.574 & 46.22  \\
        \texttt{b0l-dh} & $5$ & $5$ & 1.5 & 4.13 & 177.22 & 0.70 & 16.74 & 2.541 & 39.69  \\
        \texttt{b0l-dvh} & $5$ & $10$ & 1.5 & 4.38 & 177.22 & 0.72 & 17.22 & 2.527 & 37.10  \\
        \texttt{b0m-dvl} & $10$ & $0.1$ & 1.5 & 4.51 & 177.14 & 0.75 & 17.43 & 2.460 & 54.15  \\
        \texttt{b0m-dl}  & $10$ & $0.5$ & 1.5 & 4.76 & 177.14 & 0.74 & 18.73 & 2.435 & 48.14  \\
        \texttt{b0m-dm}  & $10$ & $1$ & 1.5 & 4.76 & 177.14 & 0.72 & 19.35 & 2.424 & 45.74  \\
        \texttt{b0m-dh} & $10$ & $5$ & 1.5 & 3.88 & 177.05 & 0.67 & 17.35 & 2.395 & 39.91  \\
        \texttt{b0m-dvh} & $10$ & $10$ & 1.5 & 3.88 & 177.05 & 0.65 & 18.00 & 2.382 & 37.56  \\
        \texttt{b0h-dvl} & $50$ & $0.1$ & 1.5 & 4.01 & 176.60 & 0.78 & 17.69 & 2.120 & 54.44  \\
        \texttt{b0h-dl} & $50$ & $0.5$ & 1.5 & 4.01 & 176.60 & 0.73 & 19.05 & 2.094 & 48.32  \\
        \texttt{b0h-dm}  & $50$ & $1$ & 1.5 & 4.01 & 176.60 & 0.71 & 19.66 & 2.084 & 46.02  \\
        \texttt{b0h-dh} & $50$ & $5$ & 1.5 & 3.01 & 176.42 & 0.66 & 16.59 & 2.054 & 39.94  \\
        \texttt{b0h-dvh} & $50$ & $10$ & 1.5 & 3.38 & 176.40 & 0.74 & 16.77 & 2.037 & 36.91  \\
        \texttt{b0vh-dvl} & $100$ & $0.1$ & 1.5 & 3.51 & 176.32 & 0.71 & 18.36 & 1.970 & 53.83  \\
        \texttt{b0vh-dl} & $100$ & $0.5$ & 1.5 & 3.63 & 176.25 & 0.77 & 17.94 & 1.947 & 48.32 \\
        \texttt{b0vh-dm} & $100$ & $1$ & 1.5 & 3.63 & 176.25 & 0.75 & 18.58 & 1.935 & 45.65 \\
        \texttt{b0vh-dh} & $100$ & $5$ & 1.5 & 3.13 & 176.11 & 0.72 & 17.06 & 1.906 & 39.79 \\
        \texttt{b0vh-dvh} & $100$ & $10$ & 1.5 & 3.13 & 176.11 & 0.70 & 17.75 & 1.892 & 37.22  \\    
     \hline
     \end{tabular}
     \tablefoot{Evolutions of the fastest PIs for different initial amplitudes of the MRI modes and the parasitic velocities. We depict, from left to right: the initial MRI magnetic field, $B_0$; the initial parasitic velocity, $v_0$; the shear parameter, $q$; the modulus and angle with respect to the radial direction of the initial wavevector, $k_{\rm h}^{\rm ini}$, that leads to saturation; its modulus and angle at saturation; the saturation time, and the amplification factor. The runs are labeled so that \texttt{vl}, \texttt{l}, \texttt{m}, \texttt{h}, \texttt{vh} after \texttt{b0} (MRI amplitude) and \texttt{d} (parasitic amplitude) stand for ``very low,'' ``low,'' ``mid,'' ``high'' and ``very high,''  respectively. }
    \label{tab:runs}
\end{table*}

\begin{table*}[h]
\caption{Same as Table~\ref{tab:runs}, but using now a shear parameter $q=1.25$ and different initial amplitudes to make a better comparison with the results from \cite{Rembiasz:2016b}.}
    \centering
    \begin{tabular}{|c|c|c|c|c|c|c|c|c|c|}
        \hline
       NAME & $ B_0[10^{-3}]$ & $v_0[10^{-3}]$ & $q$ & $k_{\rm h}^{\rm ini}$ & $\theta^{\rm ini}$ $[^{\circ}]$  & $k_{\rm h}^{\rm sat}$ & $\theta^{\rm sat}$ $[^{\circ}]$ & $t_{\rm sat} $ [Orbits] & $\mathcal{A}$  \\  
       \hline   
        \texttt{b0vl-dvl-q125} & $3.33$ & $0.112$ & 1.25 & 3.76 & 176.76 & 0.65 & 19.16 & 2.613 & 47.69  \\
        \texttt{b0vl-dl-q125} & $3.33$ & $0.370$ & 1.25 & 4.13 & 176.70 & 0.74 & 18.84 & 2.584 & 42.61  \\
        \texttt{b0vl-dm-q125}  & $3.33$ & $1.12$ & 1.25 & 4.13 & 176.70 & 0.69 & 20.22 & 2.557 & 38.24  \\
        \texttt{b0vl-dh-q125} & $3.33$ & $3.70$ & 1.25 & 4.51 & 176.66 & 0.75 & 20.40 & 2.525 & 33.73  \\
        \texttt{b0vl-dvh-q125} & $3.33$ & $11.2$ & 1.25 & 3.51 &  176.53 & 0.69 & 17.90 & 2.491 & 29.54  \\
        \texttt{b0l-dvl-q125} & $11.1$ & $0.112$ & 1.25 & 3.88 & 176.31 & 0.69 & 21.15 & 2.303 & 47.16  \\
        \texttt{b0l-dl-q125} & $11.1$ & $0.370$ & 1.25 & 3.38 & 176.19 & 0.69 & 19.16 & 2.276 & 42.54  \\
        \texttt{b0l-dm-q125} & $11.1$ & $1.12$ & 1.25 & 3.38 & 176.19 & 0.64 & 20.44 & 2.252 & 38.59  \\
        \texttt{b0l-dh-q125} & $11.1$ & $3.70$ & 1.25 & 3.88 & 176.13 & 0.74 & 20.72 & 2.216 & 33.57  \\
        \texttt{b0l-dvh-q125} & $11.1$ & $11.2$ & 1.25 & 3.01 & 175.95 & 0.68 & 18.28 & 2.183 & 29.48  \\
        \texttt{b0m-dvl-q125} & $33.3$ & $0.112$ & 1.25 & 2.76 & 175.58 & 0.67 & 18.62 & 2.026 & 47.45  \\
        \texttt{b0m-dl-q125} & $33.3$ & $0.370$ & 1.25 & 2.88 & 175.53 & 0.69 & 18.91 & 1.999 & 42.67  \\
        \texttt{b0m-dm-q125} & $33.3$ & $1.12$ & 1.25 & 3.01 & 175.47 & 0.72 & 19.27 & 1.973 & 38.51  \\
        \texttt{b0m-dh-q125} & $33.3$ & $3.70$ & 1.25 & 3.14 & 175.43 & 0.73 & 19.98 & 1.941 & 34.11  \\
        \texttt{b0m-dvh-q125} & $33.3$ & $112$ & 1.25 & 2.38 & 175.19 & 0.65 & 17.93 & 1.906 & 29.60  \\
        \texttt{b0h-dvl-q125} & $111$ & $0.112$ & 1.25 & 2.26 & 174.61 & 0.66 & 18.89 & 1.720 & 47.66 \\
        \texttt{b0h-dl-q125} & $111$ & $0.370$ & 1.25 & 2.39 & 174.59 & 0.66 & 19.93 & 1.695 & 43.21  \\
        \texttt{b0h-dm-q125} & $111$ & $1.12$ & 1.25 & 2.39 & 174.59 & 0.62 & 21.16 & 1.673 & 39.59  \\
        \texttt{b0h-dh-q125} & $111$ & $3.70$ & 1.25 & 2.64 & 174.29 & 0.80 & 19.19 & 1.639 & 34.65  \\
        \texttt{b0h-dvh-q125} & $111$ & $112$ & 1.25 & 2.01 & 173.94 & 0.70 & 17.60 & 1.600 & 29.72  \\
        \texttt{b0vh-dvl-q125} & $333$ & $0.112$ & 1.25 & 2.27 & 173.35 & 0.76 & 20.15 & 1.438 & 47.25 \\
        \texttt{b0vh-dl-q125} & $333$ & $0.370$ & 1.25 & 2.27 & 173.35 & 0.71 & 21.59 & 1.413 & 42.81  \\
        \texttt{b0vh-dm-q125} & $333$ & $1.12$ & 1.25 & 2.02 & 172.88 & 0.76 & 19.14 & 1.385 & 38.40  \\ 
        \texttt{b0vh-dh-q125} & $333$ & $3.70$ & 1.25 & 2.02 & 172.88 & 0.70 & 20.88 & 1.352 & 33.71  \\
        \texttt{b0vh-dvh-q125} & $333$ & $11.2$ & 1.25 & 2.14 & 172.63 & 0.78 & 20.71 & 1.321 & 29.77  \\ 
     \hline 
\end{tabular}
    \label{tab:runs2}
\end{table*}
\end{appendix}

\end{document}